\documentclass[11pt]{article}
\usepackage{jheppub}

\usepackage{subcaption}
\usepackage{epsfig}
\usepackage{amssymb}
\usepackage{amsfonts}
\usepackage{amsbsy}
\usepackage{float}
\usepackage{makecell}
\usepackage{amsmath}
\usepackage{dsfont}
\usepackage{bbm}
\usepackage{upgreek}
\usepackage{amssymb,amscd}
\usepackage{graphicx}
\usepackage{mathrsfs}
\usepackage{amsmath,amsthm}
\usepackage{slashed}
\usepackage{ulem,lipsum}
\usepackage{dsfont}
\usepackage{tikz}
\usepackage[utf8]{inputenc}
\newcommand{\tildelambda}{\widetilde{\lambda}}

\makeatletter
\DeclareFontFamily{OMX}{MnSymbolE}{}
\DeclareSymbolFont{MnLargeSymbols}{OMX}{MnSymbolE}{m}{n}
\SetSymbolFont{MnLargeSymbols}{bold}{OMX}{MnSymbolE}{b}{n}
\DeclareFontShape{OMX}{MnSymbolE}{m}{n}{
    <-6>  MnSymbolE5
   <6-7>  MnSymbolE6
   <7-8>  MnSymbolE7
   <8-9>  MnSymbolE8
   <9-10> MnSymbolE9
  <10-12> MnSymbolE10
  <12->   MnSymbolE12
}{}
\DeclareFontShape{OMX}{MnSymbolE}{b}{n}{
    <-6>  MnSymbolE-Bold5
   <6-7>  MnSymbolE-Bold6
   <7-8>  MnSymbolE-Bold7
   <8-9>  MnSymbolE-Bold8
   <9-10> MnSymbolE-Bold9
  <10-12> MnSymbolE-Bold10
  <12->   MnSymbolE-Bold12
}{}

\let\llangle\@undefined
\let\rrangle\@undefined
\DeclareMathDelimiter{\llangle}{\mathopen}%
                     {MnLargeSymbols}{'164}{MnLargeSymbols}{'164}
\DeclareMathDelimiter{\rrangle}{\mathclose}%
                     {MnLargeSymbols}{'171}{MnLargeSymbols}{'171}
\makeatother

\def\be{ \begin{equation} }
\def\ee{ \end{equation}}

\newcommand{\eq}[1]{\begin{align}\begin{split}#1\end{split}\end{align}}

\def\exp{{\rm exp}}

\def\Tr{{\rm \,Tr}}

\def\itwopi{\frac{i}{2\pi}}

\def\one{{\hbox{ 1\kern-.8mm l}}}

\def\CA{{\cal A}}

\def\CD {{\cal D}}

\def\CG {{\cal G}}
\def\CH {{\cal H}}

\def\CN {{\cal N}}

\def\CO {{\cal O}}
\def\CP {{\cal P}}

\def\CW {{\cal W}}

\def\CO {{\cal O}}

\def\CG {{\cal G}}
\def\CH {{\cal H}}

\def\CQ {{\cal Q}}
\def\CS {{\cal S}}
\def\CT {{\cal T}}

\def\IQ{\mathbb{Q}}
\def\IR{{\mathbb{R}}}

\def\IZ{{\mathbb{Z}}}

\def\fg{\mathfrak{g}}

\def\rmk#1{\bigskip\noindent{\bf Remark} }
\def\cnj#1{\bigskip\noindent{\bf Conjecture:} }

\def\tildeH{{\widetilde{H}}}

\def\tildeh{{\widetilde{h}}}

\def\hatY{{\widehat{Y}}}

\DeclareMathAlphabet{\mathpzc}{OT1}{pzc}{m}{it}

\def\vk{{\vec{k}}}

\def\tildea{{\tilde{a}}}

\def\tildeb{{\tilde{b}}}

\def\tildelambda{{\widetilde{\lambda}}}

\usepackage{colortbl}
\definecolor{dgreen}{rgb}{0, 0.55, 0}
\definecolor{llightyellow}{rgb}{1.0, 0.95, 0.7}
\definecolor{llightblue}{rgb}{0.7, 0.9, 1.0}
\definecolor{llightpink}{rgb}{1.0, 0.85, 0.95}
\definecolor{llightgreen}{rgb}{0.7, 1.0, 0.4}
\colorlet{lightyellow}{llightyellow!50!white}
\colorlet{lightblue}{llightblue!50!white}
\colorlet{lightgreen}{llightgreen!50!white}
\colorlet{lightpink}{llightpink!50!white}
\definecolor{azure}{rgb}{0.0, 0.5, 1.0}
\definecolor{darkblue}{rgb}{0.15,0.35,0.7}
\definecolor{reddish}{rgb}{0.65, 0.2, 0.2}
\definecolor{brandeisblue}{rgb}{0.0, 0.44, 1.0}
\definecolor{ceruleanblue}{rgb}{0.16, 0.32, 0.75}
\definecolor{indigo(dye)}{rgb}{0.0, 0.25, 0.42}
\definecolor{grey}{rgb}{0.9,0.9,0.9}
\definecolor{dgrey}{rgb}{0.3,0.3,0.3}

\usepackage[framemethod=TikZ]{mdframed} 
\usepackage{tikz-cd} 
\usetikzlibrary {shapes.geometric} 

\usetikzlibrary{arrows,snakes,shapes.arrows,decorations.markings}
     \tikzset{>=triangle 90}
     \tikzstyle{bbc}=[draw,circle,fill=black,scale=.75]
     \tikzstyle{rc}=[circle,fill=red,scale=.6]
     \tikzstyle{wc}=[draw,circle,scale=.75]
     
\usetikzlibrary{decorations.pathreplacing,calligraphy}

\tikzset{snake it/.style={decorate, decoration=snake}}

\tikzset{
	on each segment/.style={
		decorate,
		decoration={
			show path construction,
			moveto code={},
			lineto code={
				\path [#1]
				(\tikzinputsegmentfirst) -- (\tikzinputsegmentlast);
			},
			curveto code={
				\path [#1] (\tikzinputsegmentfirst)
				.. controls
				(\tikzinputsegmentsupporta) and (\tikzinputsegmentsupportb)
				..
				(\tikzinputsegmentlast);
			},
			closepath code={
				\path [#1]
				(\tikzinputsegmentfirst) -- (\tikzinputsegmentlast);
			},
		},
	},
	mid arrow/.style={postaction={decorate,decoration={
				markings,
				mark=at position .5 with {\arrow[#1]{stealth}}
	}}},
}

\tikzset{
    partial ellipse/.style args={#1:#2:#3}{
        insert path={+ (#1:#3) arc (#1:#2:#3)}
    }
}

\usetikzlibrary{decorations.markings}

\tikzset{line/.style={line width=0.25mm},
curve/.style={line,smooth,tension=1},
->-/.style={decoration={
  markings,
  mark=at position #1 with {\arrow[>=stealth]{>}}},postaction={decorate}},
-<-/.style={decoration={
  markings,
  mark=at position #1 with {\arrow[>=stealth]{<}}},postaction={decorate}},
}

\tikzset{bg/.style={opacity=.5}}

\title{A SymTFT for Continuous Symmetries}

\author[a]{T.~Daniel Brennan}
\author[a,b]{and Zhengdi Sun}

\affiliation[a]{Department of Physics, University of California San Diego}
\affiliation[b]{Mani L. Bhaumik Institute for Theoretical Physics, Department of Physics and Astronomy, University of California Los Angeles, CA 90095, USA}

\emailAdd{tbrennan@ucsd.edu}
\emailAdd{zdsun@physics.ucla.edu}

\abstract{Symmetry is a powerful tool for studying dynamics in QFT: it provides selection rules, constrains RG flows, and often simplifies analysis. Currently, our understanding is that the most general form of symmetry is described by categorical symmetries which can be realized via  Symmetry TQFTs or ``SymTFTs." In this paper, we show how the framework of the SymTFT, which is understood for discrete symmetries (i.e. finite categorical symmetries), can be generalized to continuous symmetries. In addition to demonstrating how  $U(1)$ global symmetries can be incorporated into the paradigm of the SymTFT, we apply our formalism to 
 study cubic $U(1)$ anomalies in $4d$ QFTs, describe the $\IQ/\IZ$ non-invertible chiral symmetry in $4d$ theories, and  conjecture the SymTFT for general continuous $G^{(0)}$ global symmetries. 
}

\begin{document}

\maketitle

\section{Introduction and Summary}

Symmetry is a powerful tool for studying physical processes. In general, symmetries provide selection rules for dynamical processes and can be used to constrain RG flows. Recently, the notion of symmetry in quantum theories has been expanded to include the action of all topological operators which goes beyond the notion of group-like symmetries. These topological operators, along with their braiding and fusion, are instead described by category theory and are referred to as ``generalized'' or ``categorical symmetries.'' For a review of generalized/categorical symmetries see \cite{Cordova:2022ruw,Freed:2022qnc,Gaiotto:2014kfa,Schafer-Nameki:2023jdn,Brennan:2023mmt,Bhardwaj:2023kri,Shao:2023gho} and sources therein. 

A particularly useful tool for studying the general symmetry structure of a quantum theory is the {Symmetry TQFT} or SymTFT for short 
\cite{Freed:2022qnc,Kaidi:2023maf,Kaidi:2022cpf,Kong:2020cie,Apruzzi:2021nmk,Bhardwaj:2023bbf,Apruzzi:2023uma,Bhardwaj:2023ayw,vanBeest:2022fss,Sun:2023xxv,Zhang:2023wlu,Cordova:2023bja,Antinucci:2023ezl,Cvetic:2023plv,Baume:2023kkf,Damia:2022bcd}. 
To a given $d$-dimensional QFT, we can associate a $(d+1)$-dimensional TQFT (defined by the symmetry category\footnote{
Here we will not define the symmetry category as it is relatively complicated and even the type of category differs by the dimension of the QFT/SymTFT.
}) which encodes all of the symmetries (and their anomalies \cite{Kaidi:2023maf, Sun:2023xxv, Zhang:2023wlu, Cordova:2023bja, Antinucci:2023ezl}) of the physical QFT. 
If we take our QFT on the spacetime manifold $X_d$, then the associated SymTFT  is placed on the manifold $Y_{d+1}=X_d\times [0,1]$. If we parametrize the interval (sometimes called ``the sandwich'' or ``the slab") by a coordinate $t\in [0,1]$, then the boundaries of $Y_{d+1}$ at $t=0,1$ must correspond to non-trivial boundary conditions of the SymTFT. By convention, we take $t=0$ to be the boundary associated to the QFT while the boundary at $t=1$ is a topological boundary (called the ``quiche'' boundary).
The SymTFT is comprised of a collection of topological operators which describe the possible topological symmetry operators in the QFT as well as their fusion, and linking. Their expression in the QFT is explicitly controlled by the choice of quiche boundary condition. 

\begin{figure}
\centering
    \begin{subfigure}[t]{0.5\textwidth}
    \centering
    \includegraphics[scale = 0.55]{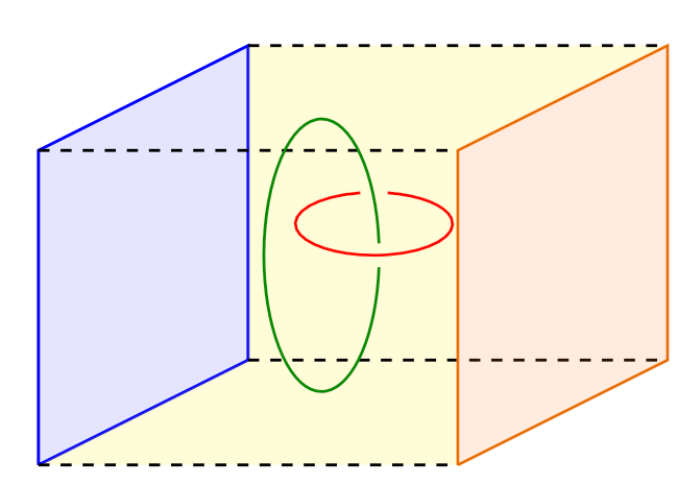}
    \caption{}
    \end{subfigure}%
    ~
    \begin{subfigure}[t]{0.5\textwidth}
    \centering
    \includegraphics[scale = 0.55]{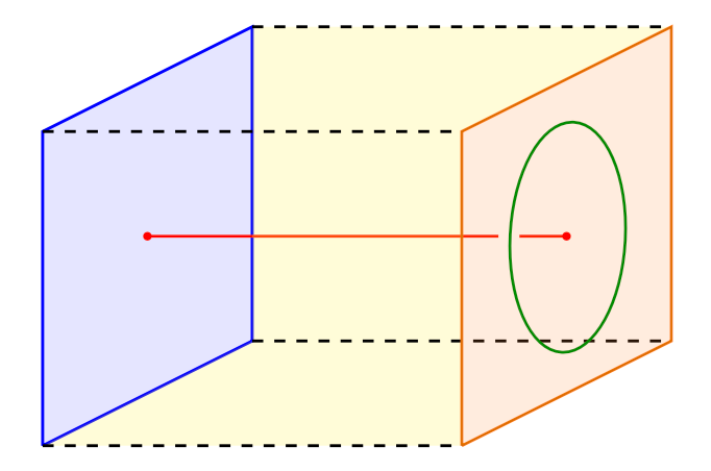}
    \caption{}
    \end{subfigure}
    
    \caption{In this figure we illustrate the idea of the SymTFT in the sandwich/slab configuration where the QFT boundary is on the left (blue) and the quiche boundary is on the right (orange). In (a) we illustrate the SymTFT as a TQFT on the interval which admits a set of topological operators with non-trivial braiding. In (b) we show some topological operators (green lines) of the SymTFT can be pushed into the boundary where they act as the topological operators that generate the symmetry, while other topological operators (red lines) can terminate on the boundary representing operators that are charged under the global symmetries.}
    \label{fig:SymTFT}
\end{figure}

See for example Figure \ref{fig:SymTFT}. The advantage of the SymTFT over the standard construction of SPT phases is that it admits a description of more general global symmetries in QFTs such as non-invertible and categorical symmetries and allows one to capture the topological manipulations via changing the symmetry boundaries.

The utility of the SymTFT is that it gives us a uniform mechanism to extract the ``topological sector'' of a QFT. Because the interval is topologically trivial, the path integral of the SymTFT on the interval identifies the topological operators on the QFT boundary with their realization on the quiche boundary. Alternatively, since the SymTFT is topological, we can dimensionally reduce along the interval, colliding the quiche boundary with the QFT boundary, thereby fixing the topological sector of the QFT by the quiche boundary conditions.  

There is also a dual picture where one quantizes the SymTFT along the interval (i.e. use $t$ as a ``time'' coordinate). In this picture, the boundary conditions correspond to states on which the topological operators of the SymTFT act and the path integral on the  interval (again being topologically trivial) computes the inner product between these states. In this picture, it is clear that fixing the quiche boundary state projects the QFT onto a particular state which realizes how the topological symmetry operators act in the QFT. 

Because of the role of the quiche boundary condition in realizing how the symmetry category of the SymTFT acts in the QFT, we can discuss the possible realizations of a particular symmetry category in terms of the SymTFT on the semi-infinite line,  $\hatY_{d+1}=X_d\times \IR_+$, independent of the QFT; much in the same way one can discuss the property of groups independent of a representation. There, we can discuss all possible topological quiche boundaries and different possible symmetry protected gapped phases that can realize a given symmetry.  

While the SymTFT is a ubiquitous tool for studying symmetries in QFT, thus far it has only been used to study finite categorical symmetries including for example finite groups, duality defects, and certain non-invertible symmetries \cite{Sun:2023xxv, Zhang:2023wlu, Cordova:2023bja, Antinucci:2023ezl,Freed:2022qnc,Kaidi:2023maf,Kaidi:2022cpf,Kong:2020cie,Apruzzi:2021nmk,Bhardwaj:2023bbf,Apruzzi:2023uma,Bhardwaj:2023ayw,vanBeest:2022fss,Damia:2022bcd}. However, in order to have a complete framework to study all symmetries, one would also like to understand how to describe continuous symmetries and their interaction with finite symmetries using the framework of the SymTFT. This is important for example in studying gapless, interacting theories that arise from spontaneously breaking a continuous global symmetry. 

\bigskip
In this paper, we will demonstrate how to one can incorporate continuous symmetries into the framework of symmetry TQFTs. Here we will give a Lagrangian formulation of these theories and perform our analysis within that framework. 
 We will primarily focus on $U(1)$ $p$-form global symmetries although we will also propose a SymTFT for $G^{(0)}$ symmetries where $G$ is a continuous non-abelian Lie group. Although one may be able to describe a $U(1)^{(p)}$ global symmetry of a QFT and its 't Hooft anomalies in terms of more traditional SPT phases, their description in using the SymTFT is more powerful because it naturally allows more general manipulations such as discrete gauging and, as we will see, captures the interplay of $U(1)^{(p)}$ symmetry with more general symmetries such as non-invertible global symmetries.

For a $d$-dimensional QFT with $U(1)^{(p)}$ global symmetry we can express the SymTFT for a $U(1)^{(p)}$ global symmetry in terms of the action:
\eq{
S_{U(1)}=\frac{i}{2\pi}\int da_{p+1}\wedge \tildeh_{d-p-1}~,
}
where $a_{p+1}$ is a $(p+1)$-form $U(1)$ gauge field and $h_{d-p-1}$ is a $\IR$-valued $(d-p-1)$-form gauge field. 
This TQFT is reminiscent of the $\IZ_N$ SymTFT which is described by a BF theory \cite{Horowitz:1989ng,Maldacena:2001ss,Banks:2010zn}, and indeed, one can restrict to the $\IZ_N^{(p)}$ subsector of the $U(1)^{(p)}$ SymTFT and reproduce the standard BF action. We can heuristically think of the $U(1)^{(p)}$ as the $\IZ_N^{(p)}$ SymTFT in the limit $N\to \infty$ where $NB_{d-p-1}\mapsto h_{d-p-1}$ and $A_{p+1}\mapsto a_{p+1}$
\eq{
S_{\IZ_N}=\frac{iN}{2\pi}\int dA_{p+1}\wedge B_{d-p-1}~\longmapsto ~S_{U(1)}=\frac{i}{2\pi}\int da_{p+1}\wedge h_{d-p-1}~,
}
which straightforwardly describes the symmetry defect operators that source flat background $U(1)^{(p)}$  connections.

Similar to the $\IZ_N$ SymTFT, the $U(1)^{(p)}$ symmetry can be described in terms of the pair of topological operators 
\eq{
W_n(\gamma)=e^{i n \oint a}\quad, \quad \CW_\alpha (\Sigma)=e^{i \alpha \oint h}~, 
}
for $n\in \IZ$ and $\alpha\in U(1)$ which have non-trivial linking.\footnote{ We are unsure what the proper categorical description should be; however we suspect that it is related to the  categories of line operators in topologically twisted $3d$ $\CN=4$ Yang-Mills theory that are described in \cite{Ballin:2023rmt,Costello:2018swh,Costello:2018fnz,Garner:2022vds}. }

The possible quiche boundary conditions are given by the familiar Dirichlet and Neumann boundary conditions which fixes either $a_{p+1}$ (diagonalizes $W_n$) or $h_{d-p-1}$ (diagonalizes $\CW_\alpha$) respectively. 
However, note that the Neumann boundary condition for this SymTFT only sums over the flat $U(1)$ connections. 
In Section \ref{sec:U(1)} we analyze this theory, its operators and boundary conditions, and demonstrate how spontaneous symmetry breaking is encoded in the SymTFT. Additionally, we discuss how one can modify the structure of the SymTFT to dynamically gauge the $U(1)$ global symmetry on the boundary by summing over all of the states in all defect Hilbert spaces.

Additionally, in Section \ref{sec:applications} we discuss several applications of the continuous SymTFT such as how anomalies are realized and how they prevent the existence of Neumann boundary conditions and how non-invertible chiral $\IQ/\IZ$ symmetry in $4d$ are realized in the $U(1)^2$ SymTFT. 

Finally, in Section \ref{sec:nonabelian} we propose a symmetry TQFT that we believe may encode the continuous, non-abelian $G^{(0)}$ global symmetries in a QFT. Our proposal is that 
\eq{
S_{G^{(0)}}=\frac{i}{2\pi}\int \Tr
\left(f_2\wedge h_{d-1}\right)~,
} 
where the trace is over the defining representation. 
Here $f_2$ is the $G^{(0)}$ field strength and $h_{d-1}$ is a $Lie[G^{(0)}]=\fg$-valued $(d-1)$-form gauge field which together transform under $G^{(0)}$ gauge transformations as 
\eq{
f_2\longmapsto g^{-1}f_2g\quad, \quad h_{d-1}\longmapsto g^{-1}(h_{d-1}+D\lambda_{d-2})g~,
}
{where $\lambda_{d-2}$ is a $\fg$-valued $(d-2)$-form transformation parameter.

In addition to the fact that this is the clear generalization of the abelian action, it is also similarly related to the topologically B-twisted $3d$ $\CN=4$ $G^{(0)}$ gauge theory \cite{Ballin:2023rmt,Costello:2018swh,Costello:2018fnz,Garner:2022vds}. The non-abelian BF theories we propose here have additionally been studied in dimension four in \cite{Cattaneo:1995tw,Cattaneo:1997eh}.

This theory contains a series of Wilson lines $W_R=\Tr_R\CP\,e^{i \oint a_1}$, admits both Dirichlet and Neumann boundary conditions, and can describe anomalies in analogy with the SymTFT for $U(1)$ global symmetry. However, these TQFTs are non-trivial and require further study as it is unclear what the full spectrum of topological operators are in this theory (due to issues with normal ordering for non-abelian Wilson-type operators of dimension greater than 1) and what becomes the $G^{(0)}$ symmetry defect operator in the QFT with Dirichlet boundary conditions.  

\bigskip
\noindent\textbf{Note Added:} 
While preparing edits for the second version of this paper, the papers \cite{Antinucci:2024zjp,Apruzzi:2024htg,Bonetti:2024cjk} were also submitted, which discuss similar ideas.

\section{SymTFT Review}
In this section, we will briefly review the idea of the SymTFT \cite{Freed:2022qnc}, taking the case of $\mathbb{Z}_N$ $0$-form symmetry as our primary example. For more details see \cite{Freed:2022qnc,Kaidi:2023maf,Kaidi:2022cpf,Kong:2020cie,Schafer-Nameki:2023jdn,Bhardwaj:2023kri,Shao:2023gho,Apruzzi:2021nmk}.

Consider a $d$-dimensional QFT $\CT$ on a spacetime manifold $X_d$. We will assume that this theory has a global symmetry structure that is determined by a collection of topological operators. Due to the standard picture of anomaly inflow, it is natural to expect that one may be able to describe these topological symmetry operators in terms of a $(d+1)$-dimensional TQFT on a manifold $Y_{d+1}$ which has a boundary component $X_d$. In this picture, the topological operators of the TQFT would become the topological symmetry operators of the QFT on the boundary, but their braiding and fusion would be determined by the behavior of the bulk operators in the TQFT. 

However, for any non-trivial symmetry, one must have a non-trivial TQFT which in general 
will have a non-trivial dependence (i.e. the Hilbert space, partition function, and etc.) on the choice of bounding manifold $Y_{d+1}$. For example, if $Y_{d+1}$ has non-trivial bulk cycles/topology away from its boundary $\partial Y_{d+1}=X_d$ (i.e. non-trivial $H_n(Y_{d+1},X_d)$), then the TQFT partition function will sum over all possible topological operators wrapping these cycles. 

The framework of the the \textit{Symmetry TQFT} or (``SymTFT" for short) indeed uses this idea, but solves the problem of choosing a $(d+1)$-dimensional manifold in a very clever way. The SymTFT gives a canonical choice of $Y_{d+1}$ by coupling the QFT on $X_d$ to a TQFT in one higher dimension on $Y_{d+1}=X_d\times[0,1]$ where  $t\in [0,1]$  parametrizes the interval where $t=0$ is the boundary on which the dynamical QFT resides. Since the interval is topologically trivial there will be no dependence on the $(d+1)$-dimensional physics except on an additional choice of boundary condition at $t=1$. Since we do not want to add additional degrees of freedom introduced into our QFT by the SymTFT, we demand that the boundary condition at $t=1$ is topological (i.e.  gapped). See Figure \ref{fig:quiche} for the setup. 
For reasons that will become clear, we will refer to this boundary as the ``quiche boundary."

\begin{figure}
\centering
    \begin{subfigure}[t]{0.5\textwidth}
        \centering
        \includegraphics[scale = 0.55]{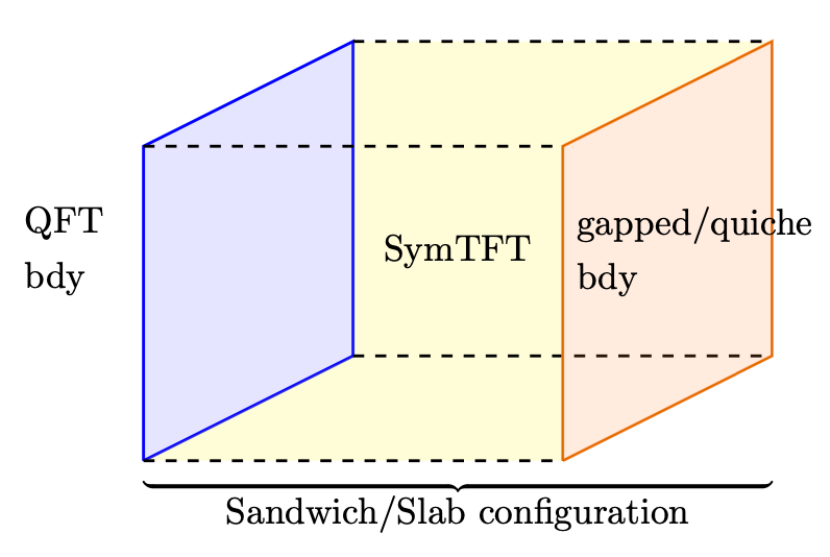}
        \caption{}
    \end{subfigure}%
    ~ 
    \begin{subfigure}[t]{0.5\textwidth}
        \centering
        \includegraphics[scale = 0.55]{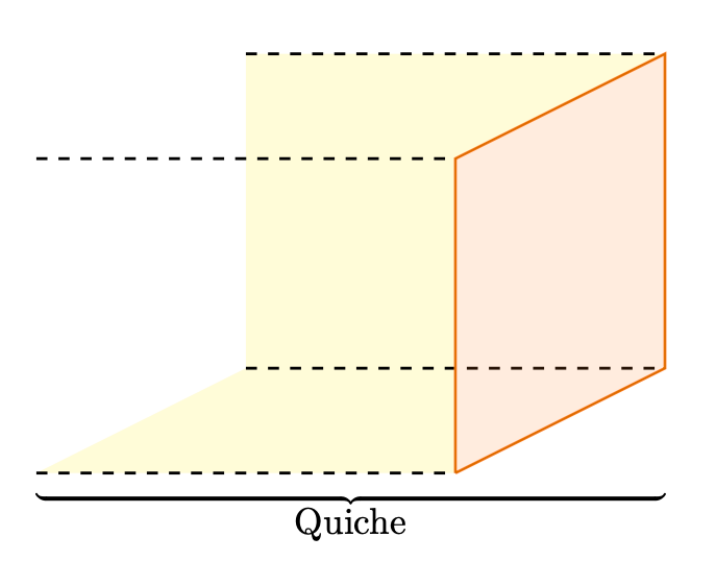}
        \caption{}
    \end{subfigure}

\caption{In this figure we illustrate the setup of the SymTFT. In (a) we show the SymTFT as a TQFT on the interval which has a gapped boundary on one end (quiche boundary) and the dynamical QFT at the other end. This configuration is sometimes called ``the sandwich" as we can collapse the interval due to the fact that the SymTFT is topological, thereby ``sandwiching" the two boundaries together. 
In (b) we show the SymTFT where we only focus on the gapped boundary. This configuration is often called ``the quiche" as the SymTFT has an open boundary. \label{fig:quiche}}
\end{figure}

This construction allows us to isolate the behavior of the topological symmetry operators of the QFT and describe them in terms of the topological operators of the $(d+1)$-dimensional SymTFT. One way to see this is the following. Since the SymTFT is topological and the interval is topologically trivial, the theory does not depend on the size of the interval. In particular, we can take the limit as the size of the interval goes to zero. In this limit, we are effectively taking the product of the topological boundary and the QFT, so the bulk operators are completely reduced to those that exist in both the QFT and the quiche boundary. 

Here, the product reproduces the path integral of the QFT in a certain phase that is determined by the quiche boundary conditions. This can be computed by either computing the partition function on the sandwich with the appropriate boundary conditions  or equivalently by taking the inner product of the QFT boundary state with the quiche boundary state. In this way, the SymTFT encodes the possible topological manipulations {one can perform on the QFT path integral in terms of the different choices of  quiche boundary conditions.}

This picture of taking the product of the QFT with the quiche boundary by reducing the interval is an intrinsic feature of the SymTFT: we can think of it as defining an action of the topological boundary of the SymTFT on the QFT (in our convention the SymTFT acts on the QFT as a right module).  Because of this action, we can think of the SymTFT with the quiche boundary as an independent object, much like how we study groups independently of their representations. This object (i.e. the SymTFT on a half-space $X_d\times \IR_-$) is often called ``the quiche."

\subsection{$\IZ_N$ SymTFT}

Let us now specify to the example of a $d$-dimensional QFT with $\IZ_N^{(0)}$ global symmetry on $X_d$.\footnote{For simplicity, we will only focus on the case of 0-form global symmetries, but the cases for general $\IZ^{(p)}_N$ symmetries will follow with straightforward modification. }
We want to couple the $d$-dimensional theory to the $(d+1)$-dimensional SymTFT on $X_d\times [0,1]$. 
Here, the SymTFT is described by the $1$-form $\IZ_N$ BF theory which has the Lagrangian:
\eq{
S=\frac{iN}{2\pi}\int da_{1}\wedge b_{d-1}~,
}
where $a_1$ and $b_{d-1}$ are $U(1)$-valued  1-form  and $(d-1)$-form gauge fields respectively. 
This theory contains $\mathbb{Z}_N$ Wilson lines of $a_{1}$ gauge field and the $\mathbb{Z}_N$ Wilson surfaces of the $b_{d-1}$ gauge field\footnote{The operators $W_{pN},\CW_{qN}$ where $p,q\in \IZ$ act as trivial operators in the TQFT because they  have trivial linking with all operators and can be absorbed by a shift of $b_{d-1}$ or $a_1$ respectively by a non-flat gauge field.}
\begin{equation}
	W_n(\gamma) = e^{i n \oint_\gamma a_{ 1}}~, \quad \CW_m(\Gamma) = e^{i m \oint_\Gamma b_{d-1}}~, \quad n,m \in \mathbb{Z}_N~.
\end{equation}
These operators have the following non-trivial braiding relation
\begin{equation}
	\langle W_n(\gamma) \CW_m(\Gamma) \rangle = e^{2\pi i \frac{mn}{N} \operatorname{Link}(\gamma,\Gamma)}~.	
\end{equation}
We additionally want to point out that  the surface operators $e^{i\alpha \oint da_{ 1}}$ and $e^{i\beta \oint dh_{d-1}}$ are topological and gauge invariant, but are trivial operators in the SymTFT.

We now want to consider the SymTFT quiche. In any (Euclidean) TQFT there is a one-to-one mapping between codimension 1 boundary conditions on $X_d$ and states in the TQFT Hilbert space quantized on $X_d$: $\CH[X_d]$. The reason is that the Euclidean signature of the TQFT allows us to quantize along the $t$-direction or along some orthogonal direction along $X_d$. This gives two equivalent descriptions of a spacetime boundary in a TQFT. 

For the $\IZ_N$ SymTFT, we can use the fact that $a_{ 1}$ and $b_{d-1}$ are canonically conjugate variables to see that there are two dual bases of orthonormal states/boundary conditions for the SymTFT: 1.) states that diagonalize the $a_{ 1}$ and 2.) states that diagonalize the $b_{d-1}$ fields. More precisely, the two classes of states diagonalize the gauge invariant operators 1.) $W_n(\gamma)$ and 2.) $\CW_n(\Gamma)$. By convention, we call these boundary conditions 1.) ``Dirichlet" denoted $|D_A\rangle$ and 2.) ``Neumann" denoted $|N_B\rangle$: 
\eq{
\text{1.) }W_n(\gamma)|D_{A}\rangle=e^{i n \oint_\gamma A}|D_{A}\rangle~, \quad 
\text{2.) }\CW_p(\Sigma)|N_{B}\rangle =e^{i n \oint_\Sigma B}|N_{B}\rangle~. 
}
Generally, we will work with the basis of Dirichlet states which we will usually write as $|A\rangle:=|D_{A}\rangle$. We will always refer to the Neumann states by $|N_B\rangle$ and will revert to the notation $|D_A\rangle$ for Dirichlet state whenever there is possible ambiguity. 

As is standard in canonical quantization, these two boundary conditions are related by a  Fourier transform: 
\eq{\label{DtoNZN}
|N_{B}\rangle=\frac{1}{\sqrt{|H^{ 1}(X_d,\mathbb{Z}_N)|}} \sum_{A\in H^{ 1}\left(X_d;\frac{2\pi}{N}\IZ_N\right)} e^{\frac{i N}{2\pi} \int A\cup B }\,|A\rangle~. 
}
Note that here we choose to normalize $A_1$ so that it is a $\IZ_N\subset U(1)$ gauge field -- matching most of the  discussion in our paper. 

Similarly, the Dirichlet boundary condition can be constructed from the Neumann boundary condition by inverse Fourier transform:
\eq{
|A\rangle=\frac{1}{\sqrt{|H^{d-1}(X_d,\mathbb{Z}_N)|}}\sum_{B \in H^{d-1}\left(X_d;\frac{2\pi}{N}\IZ_N\right)}e^{-\frac{iN}{2\pi}\int A\cup B }~|N_{B}\rangle~.
}
This procedure which allows us to go back-and-forth between Dirichlet and Neumann boundary conditions is formally gauging the associated $\IZ_N^{(0)}$ or $\IZ_N^{(d-2)}$  global symmetry as appropriate on the boundary. This procedure is often referred to as a type of \textit{condensation} as we can implement these gaugings by summing over all possible boundary insertions of the $b$-surfaces or $a$-lines respectively.
Thus in the two cases, we condense the $b$-surface operators to go from Dirichlet to Neumann (since they implement the gauge transformations for the $a_1$ gauge field) or $a$-line operators to go from Neumann to Dirichlet respectively.

To better understand the notion of condensation, let us first consider the Dirichlet boundary condition. Here, the $a$-line operators are diagonalized by the states $|A\rangle$. On the other hand, the $b$-surface operators act non-trivially on the Dirichlet states by shifting $A$ by a flat $\IZ_N$ gauge field since the $b$-surface operators source a flat background gauge field for $a$. 

A closely related construction of the  Dirichlet states in $\CH[X_d]$ are the states in the defect Hilbert space $\CH_{W_n}[X_d]$. Here we construct the defect Hilbert space by inserting a Wilson line $W_n(\gamma)$ so that it stretches along the $t$-direction and intersects $X_d$ along at a point $x\in X_d$ and quantizing the theory on $X_d$ in this background. This Hilbert space is spanned by Dirichlet states which again diagonalize the $a$-lines. However, due to the fact that the $W_n(\gamma)$ have non-trivial linking with the $\CW_p(\Sigma)$, we see that the associated Neumann states are all trivial. This should come as no surprise because going from Dirichlet to Neumann is accomplished by gauging a symmetry under which all of the states in $\CH_{W_n}[X_d]$ are charged. 

Since the Neumann boundary condition $|N_B\rangle$ is the analogous Dirichlet boundary condition for the $b$-surface operators $(\CW_p)$, we can similarly define the defect Hilbert space $\CH_{\CW_p}[X_d]$ where we have inserted a bulk $\CW_p(\Sigma)$ operator that stretches along the time direction so that $\Sigma$ intersects $X_d$ along a $(d-p-2)$-manifold $\sigma$. For similar reasons, the $\CH_{\CW_p}[X_d]$ does not admit boundary conditions which diagonalize the $W_n(\gamma)$ operators since this would require gauging the $\IZ_N^{(d-2)}$ global symmetry under which all states in $\CH_{\CW_p}[X_d]$ are charged. 

Often, we will not differentiate between the Dirichlet states of $\CH_{W_n}[X_d]$ and $\CH[X_d]$ or the  Neumann states of $\CH_{\CW_p}[X_d]$ and $\CH[X_d]$. Rather we will think of the states of the defect Hilbert space $|A\rangle_{W_n}\in \CH_{W_n}[X_d]$ as constructed from $|A\rangle\in \CH[X_d]$  and $|N_B\rangle_{\CW_p}\in \CH_{\CW_p}[X_d]$ as constructed from $|N_B\rangle \in \CH[X_d]$ which  we ``dress" with (or really intersect with) a bulk $W_n(\gamma)$ or $\CW_p(\Sigma)$ operator as appropriate. 
With this viewpoint, we can say that if we start with a Dirichlet boundary condition $|A\rangle$, we can end a $\CW_n$ line operator on the boundary. However, condensing the $\CW_n$ operators (i.e. gauging the $\IZ_N^{(0)}$ symmetry on the boundary) so that when we pass from $|A\rangle\longmapsto |N_B\rangle$  ending the $W_n$ operators are prevented on the boundary.

\bigskip
Now that we have discussed the  $\IZ_N^{(0)}$ SymTFT, we would like to discuss how the $\IZ_N^{(0)}$ quiche acts on a QFT with $\IZ_N^{(0)}$ global symmetry. Because we are considering a theory with a group-like global symmetry, we know explicitly how to couple the partition function to a background gauge field: $Z_{\CT }[A_{1}]$. Because of this, we can also gauge the symmetry to arrive at the theory $\CT/\IZ_N$ by summing over the $\IZ_N^{(0)}$ background gauge fields:
\eq{
Z_{\CT/\IZ_N 
}[B_{d-1}]=\frac{1}{\sqrt{|H^1(X_d;\mathbb{Z}_N)|}}\sum_{A_{ 1}\in H^{ 1}\left(X_d;\frac{2\pi}{N}\IZ_N\right)} e^{\frac{iN}{2\pi}\int A_{ 1}\cup B_{d-1}}~Z_{\CT 
}[A_{ 1}]~,
}
where we have included a background gauge field $B_{d-1}$ for the quantum/dual $\IZ_N^{(d-2)}$ global symmetry. The SymTFT allows us to unify both of these in terms of a state representing the  boundary QFT which is given by
\eq{
\langle {\rm QFT}
|=\sum_{A \in H^{1}\left(X_d;\frac{2\pi}{N}\IZ_N\right)} Z_{\CT 
}[A]~\langle A|~.
}
Additionally, we can also present the state in terms of the Neumann boundary conditions by 
\eq{
\langle {\rm QFT} 
|=\sum_{B\in H^{d-1}\left(X_d;\frac{2\pi}{N}\IZ_N\right)}{Z}_{\CT/\IZ_N 
}[B]~\langle N_B|~.
}
We can then realize the background and dynamically gauged theories by sandwiching the SymTFT quiche with Dirichlet and Neumann boundary conditions respectively. In terms of the Dirichlet presentation of $\langle {\rm QFT} 
|$, the inner product is given by 
\eq{
\langle {\rm QFT} 
|A\rangle&=Z_{\CT 
}[A]~,\\
\langle {\rm QFT} 
|N_B\rangle&=\sum_{A\in H^{1}\left(X_d;\frac{2\pi}{N}\IZ_N\right)} \frac{e^{\frac{i N}{2\pi} \int A\cup B}\langle {\rm QFT} 
|A\rangle}{\sqrt{|H^{1}(X_d,\mathbb{Z}_N)|}} \\&=\sum_{A\in H^{1}\left(X_d;\frac{2\pi}{N}\IZ_N\right)} \frac{e^{\frac{iN}{2\pi}\int A\cup B}~Z_{\CT}[A]}{\sqrt{|H^{1}(X_d,\mathbb{Z}_N)|}}=Z_{\CT/\IZ_N}[B]~.
}

Here,  the defect Hilbert spaces $\CH_{W_n},\CH_{\CW_p}$ also have a natural interpretation. In particular, we can end the $a/b$-Wilson operators on  charged operators in the QFT. These two perspectives are more natural in the Dirichlet/Neumann presentation of $\langle \text{QFT}|$ respectively in which case we can elevate each term in the sum
\eq{
&Z_{\CT}[A]~\langle A|~\longmapsto ~\langle \CO_p(x)\rangle_A~\langle A|_{W_p}~,\\
&Z_{\CT/\IZ_N}[B]~\langle N_B|~\longmapsto~ \langle \widetilde\CO_{p'}(\sigma)\rangle_B~\langle N_B|_{\CW_{p^\prime}}~,
}
where here $\CO_p(x)$ carries charge $p$ under $\IZ_N^{(0)}$ and $\widetilde\CO_{p'}(\sigma)$ carries charge $p'$ under the quantum/dual symmetry $\IZ_N^{(d-2)}$. 

Note that whether or not there exists a gauge invariant operator that is charged under the $\IZ_N^{(0)}$ or $\IZ_N^{(d-2)}$ global symmetry is dependent on the realization of the $\IZ_N^{(0)}$ global symmetry in the QFT. For the case of $\CO_p(x)$, the operator is only charged under a global symmetry in the case where we do not gauge $\IZ_N^{(0)}$. In the gauged case, $\CO_p(x)$ is not a gauge invariant local operator and does not constitute a good operator in our theory -- rather it must be dressed by a Wilson line $W_p(\gamma)$ where $\partial \gamma= x$ in $X_d$. For the case of $\widetilde\CO_p(\sigma)$, the operator is only charged under a global symmetry when we gauge $\IZ_N^{(0)}$. In the ungauged case, $\widetilde\CO_p$ is not a well defined operator but rather must also be dressed by a $\CW_p(\Sigma)$ surface operator where $\partial \Sigma=\sigma$.

We can see this from inserting corresponding $W_p(\gamma)$ and $\CW_p(\Sigma)$ in the SymTFT and then reducing along the interval. In the case of the Neumann boundary condition on the quiche boundary, the $\IZ_N^{(0)}$ global symmetry is not gauged. 
Here, the $W_p(\gamma)$ can end on the QFT boundary but not on the quiche boundary. See Figure \ref{fig:NeumannWilsonLine}. Rather, on the quiche boundary, the $W_p(\gamma)$ operator must be continued by a boundary $W_p(\gamma)$ operator, reflecting the fact that the operator $\CO_p(\gamma)$ is not a gauge invariant operator. On the other hand, the $\CW_p(\Sigma)$ operators can end on the the QFT and Neumann boundary state where they source an operator that is charged under the dual quantum $\IZ_N^{(d-2)}$ -- i.e. a vortex-type operator. 

Now consider reducing the SymTFT along the interval with the Dirichlet boundary condition so that the $\IZ_N^{(0)}$ global symmetry is not gauged in the QFT. Now, we can end the $W_p(\gamma)$ Wilson line on both the QFT and quiche boundary in which case we can interpret the bulk $W_p(\gamma)$ as enforcing the transformation properties of $\CO_p(x)$ in the QFT under $\IZ_N^{(0)}$ global transformations that are enacted by bulk $\CW_p(\Sigma)$ operators. In the case of the Dirichlet boundary condition, we can also end the $\CW_p(\Sigma)$ operator on the QFT boundary, but in the quiche boundary, it must be continued by a boundary $\CW_p$ operator. This reflects the fact that $\widetilde\CO_p(\sigma)$ is not a well defined operator and must be attached to a $\CW_p(\Sigma)$ operator as with $\CO_p$ for Dirichlet boundary conditions.

\begin{figure}
    \centering
    \includegraphics[scale = 0.55]{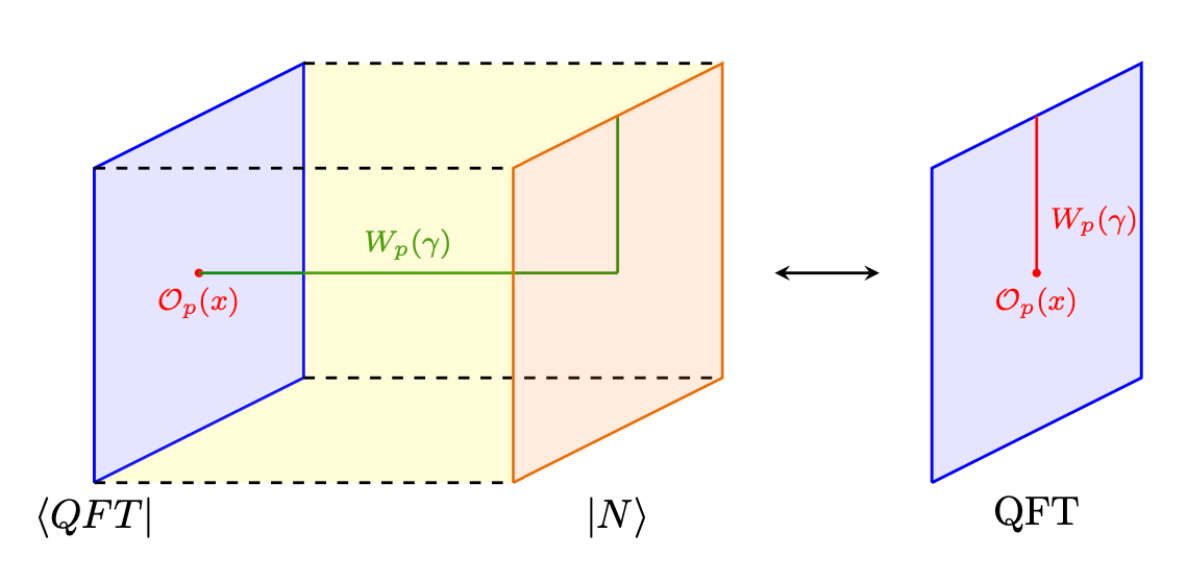}
    \caption{In this figure we illustrate how the operator $\CO_p(x)$ in the QFT requires dressing with a Wilson line $W_p(\gamma)$ for the case of Neumann boundary conditions.}
    \label{fig:NeumannWilsonLine}
\end{figure}
\subsection{Reducing to $\IZ_M\subset \IZ_N$ SymTFT}

One feature which will be important for our discussion of the $U(1)$ SymTFT is how we can reduce the SymTFT from $\IZ_N\to \IZ_M$ where $M$ divides $N$. This reduction can be realized in two complimentary ways. 

First, let us consider taking the action for the $\IZ_N^{(0)}$ SymTFT:
\eq{
S=\frac{i N}{2\pi}\int da_1\wedge b_{d-1}~,
}
and decompose $N=nM$. We can reduce to the $\IZ_M^{(0)}$ SymTFT if we restrict
\eq{
B_{d-1}=n\, b_{d-1}~. 
}
If we plug this restriction directly into the action, we find 
\eq{
S=\frac{inM}{2\pi}\int da_1\wedge b_{d-1}~\longmapsto \frac{iM}{2\pi}\int da_1\wedge B_{d-1}~,
}
which indeed describes the $\IZ_M^{(0)}$ SymTFT. This corresponds to restricting the set of  operators 
\eq{
W_p=e^{ip \oint a_1}\quad, \quad \widetilde\CW_{qn}=e^{i qn  \oint b_{d-1}}\quad, \quad p,\,q=0,1,...,M-1~. 
}
We can also think of this reduction from $\IZ_N\mapsto \IZ_M$ as a projection which can be enacted by  gauging the $\IZ_n^{(d-1)}$ subgroup which is generated by $W_M=e^{i M \oint a_1}$. Here we see that this gauging will restrict the operators $W_p$ for $p=0,...,M-1$ and project out the operators that have non-trivial linking with it: $\CW_q$ where $q\notin n\IZ$. 

There is an alternative reduction of the $\IZ_N$ SymTFT to the $\IZ_M$ SymTFT. Instead of gauging the $\IZ_n^{(d-1)}$ global symmetry of the $\IZ_N$ BF theory, we can instead gauge the $\IZ_n^{(1)}$ global symmetry. This sums over all insertions of the operators $\CW_{Mq}$. This reduces the set of non-trivial line operators to $\CW_q$ where $q=0,...,M-1$ and projects out the Wilson lines except those of the form $W_{np}$. At the level of the Lagrangian, this is equivalent to presenting the SymTFT as 
\eq{
S_{\IZ_N}^\prime=\frac{iN}{2\pi}\int a_1\wedge db_{d-1}~,
}
and restricting $A_1=na_1$ 
so that 
\eq{
S_{\IZ_N}^\prime=\frac{inM}{2\pi}\int a_1\wedge db_{d-1}~\longmapsto ~\frac{iM}{2\pi}\int A_1\wedge db_{d-1}~. 
}
 These two reductions describe similar physics and simply correspond to a choice of operators that generate the $\IZ_M$ global symmetry. 

\bigskip
More generically, it is possible to decompose $\mathbb{Z}_{NM}^{(0)}$-SymTFT into a coupled $\mathbb{Z}_M^{(0)}$- and $\mathbb{Z}_N^{(0)}$-SymTFT. This coupling is determined by whether or not $\IZ_{NM}$ splits as a direct product of $\IZ_N\times\IZ_M$ or not. This depends on whether or not ${\rm gcd}(N,M)$ is non-trivial. For our following discussion we will use the presentation of $\IZ_N$ discrete gauge theory in terms of discrete cohomology.

In the case where $\operatorname{gcd}(M,N) = 1$, $\IZ_{NM}=\IZ_N\times\IZ_M$ and the $\mathbb{Z}_{NM}^{(0)}$-SymTFT trivially factorizes into a $\mathbb{Z}_M^{(0)}$-SymTFT and a $\mathbb{Z}_N^{(0)}$-SymTFT.  
This can be seen by starting with the $\IZ_{NM}^{(0)}$-SymTFT
\eq{
S_{NM}=\frac{2\pi i}{NM}\int A_1\cup \delta B_{d-1}~,
}
where the fields are discrete co-chains $B_{d-1}\in C^{d-1}(M;\IZ_{NM}),~ A_1\in C^1(M;\IZ_{NM})$. Since  $\gcd(M,N) = 1$, there exist $p,q\in \IZ$ such that 
\begin{equation}
    pM + qN = 1~,
\end{equation}
which allows us to decompose 
\begin{equation}\label{decompZNtrivial}
    A_1 = qN a_1^{(M)} + pM a_1^{(N)}\quad, \quad B_{d-1} = N b_{d-1}^{(M)} + M b_{d-1}^{(N)}~.
\end{equation}
To see that this is a ``faithful" change of variable, notice that $\oint a_1^{(M)} = \oint a_1^{(N)} = 1$ corresponds to $\oint A_1 = 1$ and $\oint b_{d-1}^{(M)} = q,~ \oint b_{d-1}^{(N)} = p$ corresponds to $\oint B_{d-1} = 1$; thereby generating the entire field space. If we then plug this decomposition into the action we find
\eq{
S=\frac{2\pi i\, pM}{N}\oint a_1^{(N)}\cup \delta b_{d-1}^{(N)}+\frac{2\pi i \, qN}{M}\oint a_1^{(M)}\cup \delta b_{d-1}^{(M)}~,
 }
 which can be brought to the form 
 \eq{
S=\frac{2\pi i\, }{N}\oint a_1^{(N)}\cup \delta b_{d-1}^{(N)}+\frac{2\pi i \, }{M}\oint a_1^{(M)}\cup \delta b_{d-1}^{(M)}~,
 }
by adding the integral counter terms 
\eq{
S_{c.t.}=2\pi i q \oint a_1^{(N)}\cup \delta b_{d-1}^{(N)} +2\pi i p\oint a_1^{(M)}\cup \delta b_{d-1}^{(M)}~.
}

Indeed, the spectrum of operators can be matched between the $\IZ_{NM}^{(0)}$-SymTFT and that of the product SymTFT. Denoting $(W_1, \CW_1), (W_1',\CW_1')$ as the generators of the spectrum of topological operators of the $\mathbb{Z}_M^{(0)}$-SymTFT and the $\mathbb{Z}_N^{(0)}$-SymTFT respectively, then 
\eq{
\left(W_q W_p'~,~ \CW_1 \CW_1'\right):=\left((W_1)^q (W_1^\prime)^p~,~\CW_1\CW_1^\prime
\right)
}
generate the topological operators of the $\mathbb{Z}_{MN}^{(0)}$-SymTFT. 

When  $M,N$ are not coprime, $\IZ_{NM}$ is more generally an extension of $\IZ_N$ by $\IZ_M$. Due to the factorization when gcd($M,N)=1$, it suffices to demonstrate how to  factorize the $\IZ_{N^{p+q}}^{(0)}$-SymTFT into $\IZ_{N^p}$ and $\IZ_{N^q}$ components. In this case, the decomposition in \eqref{decompZNtrivial} is modified to 
\eq{
A_1=N^q a_1+\tildea_1\quad, \quad B_{d-1}=N^p \tildeb_{d-1}+b_{d-1}~,
}
where $a_1,b_{d-1}$ are $\IZ_{N^p}$-valued gauge fields and $\tildea_1,\tildeb_{d-1}$ are $\IZ_{N^q}$-valued gauge fields. This decomposition is supplemented by the additional shifts in the $\IZ_{N^{p+q}}$  lift:
\eq{\label{ZNMtrans}
&a_1\longmapsto a_1+N^p \lambda_1-\tilde\lambda_1\quad~\,, \quad \tildea_1\longmapsto \tildea_1+N^q\tilde\lambda_1~,\\
&b_{d-1}\longmapsto b_{d-1}+N^p\Lambda_{d-1}\quad, \quad \tildeb_{d-1}\longmapsto \tildeb_{d-1}+N^q \widetilde\Lambda_{d-1}-\Lambda_{d-1}~.
}
 Plugging this into the action, we get 
\eq{\label{ZNpq}
S=\frac{2\pi i }{N^p}\int a_1\cup \delta b_{d-1}+\frac{2\pi i }{N^q}\int \tildea_1\cup \delta \tildeb_{d-1}+\frac{2\pi i }{N^{p+q}}\int \tildea_1\cup \delta b_{d-1}~,
}
up to integral terms. Here the mixed term can be interpreted as a sort of ``mixed anomaly'' which requires the extension of the symmetry transformations above \eqref{ZNMtrans}. 

We can additionally check that the action in \eqref{ZNpq} realizes the operator spectrum for $\IZ_{N^{p+q}}^{(0)}$-SymTFT. Here, because of the gauge transformations in \eqref{ZNMtrans} that are necessary for the action to be invariant under the $\IZ_{N^p}^{(0)}$ gauge transformations, the $b_{d-1}$-surfaces must be of the form 
\eq{
\CW_k={\rm exp}\left\{\frac{2\pi i \,k}{N^{p+q}}\oint \left(N^p \tildeb_{d-1}+b_{d-1}\right)\right\}\quad, \quad k=0,...,N^{p+q}-1~,
}
and the Wilson lines must be of the form 
\eq{
W_p={\rm exp}\left\{\frac{2\pi i\,p}{N^{p+q}} \oint \left(N^q a_1+\tildea_1\right)\right\}\quad, \quad q=0,...,N^{p+q}-1~,
}
which together generate the topological operators of the $\IZ_{N^{p+q}}^{(0)}$-SymTFT. Here the quantization of operators has additional factors of $2\pi/N^{p+q}$ due to the fact that we are working with the integral-valued fields.

\subsection{Anomalies of $\IZ_N^{(0)}$  in the SymTFT}\label{sec:ZN_anomaly}

One powerful feature of the SymTFT is that it provides a way to encode  both the global symmetries of a QFT and their anomalies \cite{Kaidi:2023maf, Zhang:2023wlu, Cordova:2023bja, Antinucci:2023ezl}. Although we do not say that the SymTFT nor the symmetries are innately anomalous, any realization of the symmetry in a QFT or conversely an action of the SymTFT (thought of as the TQFT with a quiche boundary) on a QFT will be anomalous. 

Let us illustrate how these anomalies can be realized in the case of the $\IZ_N^{(0)}$ SymTFT with an example. In $4d$ QFTs with a $\IZ_N^{(0)}$ global symmetry, there is a unique, purely $\IZ_N^{(0)}$ anomaly which can be given by the $5d$ SPT phase:\footnote{In terms of discrete cohomology elements, this anomaly is given by 
\eq{
\CA=\frac{2\pi i }{6N}\kappa\int a_1\cup \beta(a_1)\cup \beta(a_1)~. 
}
}
\eq{
\CA=\frac{i\,\kappa}{24\pi^2}\int A_1\wedge dA_1\wedge dA_1~,
}
where $A_1$ is the integral lift (i.e. $U(1)$ representative) of a $\IZ_N$ gauge field which is normalized 
\eq{
e^{i \oint A_1}=e^{\frac{2\pi i n}{N}}~,\quad n\in \IZ~.
}
In the SymTFT, this anomaly is incorporated by adding a corresponding Chern-Simons term
\eq{\label{ZNSPT}
S_{\rm SymTFT}=\frac{iN}{2\pi}\int da_1\wedge b_3+\frac{i\kappa}{24\pi^2}\int a_1\wedge da_1\wedge da_1~.
} 
One of the well known features of anomalies is that they prevent the gauging of corresponding symmetry. In the context of the SymTFT, the Chern-Simons term obstructs the existence of the Neumann boundary condition. We can see this as follows.\footnote{
In the SymTFT literature \cite{Kaidi:2022cpf,Zhang:2023wlu,Kaidi:2023maf,Cordova:2023bja,Antinucci:2023ezl}, the anomaly is said to obstruct the existence of a ``fiber functor." Physically, this is the existence of a pair of boundary conditions which are ``orthogonal" in phase space. In other words, there are  no pair of boundary conditions that we can impose on the interval so that the path integral describes the trivially gapped phase.   In terms of the boundary QFT, this is the statement that an anomaly  obstructs the theory from flowing in the IR to a trivially gapped phase. 
}

First let us consider the theory with the action in \eqref{ZNSPT}. Adding the Chern-Simons term has the effect of shifting the equations of motion:
\eq{
N\frac{da}{2\pi}=0\quad, \quad \frac{N db_3}{2\pi}+\frac{\kappa}{8\pi^2}da_1\wedge da_1=0~.
}
Because of this, the Wilson line operator $e^{i n \oint a_1}$ is still topological, and to see that the $b$-surface is topological, we must use the fact that equations of motion imply $da=0$.

The fact that the anomaly prevents Neumann boundary conditions can be seen directly from studying these  operators. The Chern-Simons term in the action above can be interpreted as giving the $\CW_p$ operator a non-trivial expectation value
\eq{
\langle \CW_p(\Sigma)\rangle=e^{\frac{2\pi i}{N^3}\kappa p^3~{\rm Link}(\Sigma,\Sigma,\Sigma)}~,
}
where here the Link is given by the triple self-intersection number \cite{Zhang:2021ycl,Kaidi:2023maf,Putrov:2016qdo}.  Because of this,  condensing the $\CW_p$ operators in an attempt to construct the Neumann state from the Dirichlet state as in \eqref{DtoNZN} will lead to the empty state: $|A\rangle\mapsto 0$. In this way, the anomaly prevents the Neuamann boundary state.

We can also solve for the possible boundary conditions by studying the Lagrangian: they are given by the Lagrangian subspaces of phase space so that  the boundary contribution to the variation of the action vanishes. The boundary variation can be computed directly as:
\eq{
\delta S_{\rm SymTFT}=\frac{iN}{2\pi}\int_{X_d} \delta a_1\wedge \left(b_3+\frac{i\kappa}{6\pi N}a_1\wedge da_1\right)=0~.
}

Here, the boundary conditions can be reduced to solving:
\eq{
1.) ~\delta a_1\big{|}_{X_d}=0\quad {\rm or} \quad2.)~ \frac{N b_3}{2\pi}+\frac{\kappa}{ 3(2\pi)^2}a_1\wedge da_1\big{|}_{X_d}=0~. 
} 
Here, the first condition is the standard Dirichlet boundary condition. The second boundary condition, is the would-be Neumann boundary condition; however, there are several problems with 2.). First, the boundary conditions are not compatible with the bulk equations of motion. Since the boundary conditions are not compatible with the bulk equations of motion (in addition to not being gauge invariant), the space of solutions to the boundary conditions intersects the bulk phase space transversely except for where $a_1\wedge da_1=0$ and $b_3=0$. These restrictions are over determined -- they do not form a Lagrangian subspace of phase space -- and hence do not form good boundary conditions.\footnote{
Technically, one could consider the theory for which $a_1\wedge da_1=0$, however it is not usually what we mean by $\IZ_N$ BF theory (it would require some additional interaction or restriction on the path integral) and indeed would correspond to a strange global symmetry for which we only allow ourselves to couple to $\IZ_N$ bundles with this extra constraint that trivializes the putative anomaly. 
} Indeed, if there was a Neumann state that was constructed in this way, we would be able to trivialize the SymTFT (which corresponds to the existence of a trivially gapped phase) by considering the sandwich between the Dirichlet and Neumann state. However, it is well known that these anomalies obstruct the existence of a trivially gapped phase.

\section{SymTFT for $U(1)^{(0)}$ Symmetry}
\label{sec:U(1)}

{
In this section we discuss the SymTFT for describing $U(1)$ global symmetries.
Here, we first  present the SymTFT and study its operator content on a closed manifold $Y_{d+1}$ and then  study the canonical quantization of the theory on $X_d \times \mathbb{R}_t$. Next we consider the SymTFT on the quiche configuration where we describe its possible gapped boundaries and the behavior of the bulk operators on the boundary. Using this, we then describe how the SymTFT couples to a QFT on the interval and discuss the behavior of the $U(1)^{(0)}$ symmetry and the operators of the SymTFT in the QFT. 

We then discuss how different IR phases of a QFT with $U(1)$ global symmetry are realized in the SymTFT and how to realize different global structures of the $U(1)^{(0)}$ symmetry. Finally, we conclude the section with a discussion of how the SymTFT can be used to couple the QFT to non-flat connections and we additionally comment on the dynamical gauging of the $U(1)$ symmetry.}

\bigskip
The $(d+1)$-dimensional SymTFT for a $U(1)^{(p)}$ global symmetry in a $d$-dimensional QFT is described the action
\begin{equation}\label{eq:SymTFT_action}
    S = \frac{i}{2\pi} \int_{Y_{d+1}} da_{p+1} \wedge \widetilde{h}_{d-p-1} ~,
\end{equation}
where $a_p$ is a $p$-form $U(1)$ gauge field and $\widetilde{h}_{d-1}$ is a $(d-p-1)$-form $\mathbb{R}$ gauge field. For simplicity, we will focus on the case where $p = 0$ for the rest of this section, and drop the subscripts denoting the rank of the form. It is straightforward to generalize our discussion to the case with generic $p$.

Let us begin by studying the topological operators in the theory. From $a_1$, we can construct the Wilson line
\begin{equation}
    W_n(\gamma) = e^{in \oint_\gamma a} ~, \quad n \in \mathbb{Z} ~.
\end{equation}
Similarly, one can construct the surface operator from $\widetilde{h}_{d-1}$:
\begin{equation}
    \CQ(\Gamma) = \oint_\Gamma \widetilde{h} ~,
\end{equation}
which is gauge invariant as $\widetilde{h}_{d-1}$ is a $\mathbb{R}$ gauge field. It is convenient to introduce the Wilson type surface operator of the form
\begin{equation}
    \CW_\alpha(\Gamma) = e^{i\alpha \oint_{\Gamma} \widetilde{h}} ~, \quad \alpha \in [0,1) ~,
\end{equation}
which has non-trivial braiding with the operator $W_n(\gamma)$:
\begin{equation}
    \langle W_n(\gamma) \CW_\alpha(\Gamma) \rangle  = e^{2\pi i n \alpha \operatorname{Link}(\gamma,\Gamma)} ~.
\end{equation}
Here, $\alpha$ effectively takes value in $[0,1)$ because the flux sum over the $\frac{da_1}{2\pi}$ forces $\CQ(\Gamma)$ to be valued in $2\pi \mathbb{Z}$, thus $\CW_n(\Gamma)$ where $n \in \mathbb{Z}$ should be identified as the identity operator as we will show momentarily. 

\subsection{Canonical Quantization}

In order to study the SymTFT placed on $X_d \times [0,1]$ where $X_d$ is a compact $d$-dim manifold, one must understand the boundary conditions of the SymTFT. 
As is standard for TQFTs, the topological boundary conditions can be described in terms of the states of the TQFT where we canonically quantize along the same manifold. Here we will perform this canonical quantization to derive the allowed boundary conditions. 

For simplicity, we will assume $H^2(X_d,\mathbb{Z})$ is torsion free (or equivalently, there is no torsion $1$-cycle in $X_d$) throughout the paper. A boundary condition is specified by a state in the Hilbert space quantized on $X_d$, which we now study following \cite{Elitzur:1989nr}. For simplicity, we will take $X_d = T^d$; but the result generalizes to $X_d$ with no torsion $1$-cycle straightforwardly. For this, consider placing the theory on $X_d \times \mathbb{R}_t$, and rewrite the action as
\begin{equation}
    S = \frac{i}{2\pi} \int_{-\infty}^{\infty} dt \int_{X_d} \dot{\underline{a}} \wedge \underline{\widetilde{h}} + a_t \underline{d}\underline{\tildeh} + \underline{da} \wedge \tildeh_t ~, 
\end{equation}
where we have decomposed any $n$-form $w = \underline{w} + dt \wedge w_t$ into a $n$-form $\underline{w}$ and a $(n-1)$-form $w_t$ on $X_d$, and we use $\underline{d}$ to denote the exterior derivative on $X_d$. We also suppress the subscript denoting the degree of the forms to simplify the equations. 

We immediately see that $h_t$ and $a_t$ are Lagrangian multipliers enforcing $\underline{da} = \underline{d\tildeh} = 0$. Together with the gauge transformation of the $\underline{a}$ and $\underline{\tildeh}$, we learn that the classical phase space containing flat $U(1)$ connections $\underline{a}$ and flat $\mathbb{R}$ connections $\underline{\tildeh}$ modulo gauge transformations. It is then convenient to introduce operators
\begin{equation}
\begin{aligned}
    & X_\gamma = \oint_\gamma \underline{a}, \quad \gamma \in H_1(T^d,\mathbb{Z}) \cong \mathbb{Z}^d~, \\
    & \CQ_\Gamma = \oint_\Gamma \underline{\tildeh}, \quad \Gamma \in H_{d-1}(T^d,\mathbb{Z}) \cong \mathbb{Z}^d ~. \\
\end{aligned}
\end{equation}
Notice that while $\CQ_\Gamma$ is gauge invariant, $X_\gamma$ is not and the large gauge transformation of $\underline{a}$ will shift $X_\gamma \rightarrow X_\gamma + 2\pi n$ where $n \in \mathbb{Z}$. To proceed, we now take the basis $\{\gamma_i\}_{i=1}^d$ and $\{\Gamma_j\}_{j=1}^d$ for $H_1(X_d, \mathbb{Z})$ and $H_{d-1}(X_d, \mathbb{Z})$ such that their intersection numbers satisfy
\begin{equation}
    \#(\gamma_i \cap \Gamma_j) = \delta_{ij}~.
\end{equation}
Let us denote the corresponding operators  $X_{\gamma_i},\CQ_{\Gamma_j}$ as $X_i$ and $\CQ_j$. Due to the intersection of $\gamma_i,\Gamma_j$, we find the commutation relations 
\begin{equation}
    [X_i, \CQ_j] =  2\pi i \delta_{ij} ~.
\end{equation}
The Hilbert space can be constructed by viewing the operator $X_i$ as the coordinate and $\CQ_i$ as the momentum. Notice the large gauge transformation which forces $X_i \sim X_i + 2\pi$ means the system is a particle on a ring, therefore the eigenvalues of the momentum operator $\CQ_i$ must be quantized. There are two complete orthonormal bases of the Hilbert space. One basis (Dirichlet) diagonalizes the Wilson line operators $e^{i \vk\cdot \vec{X}}$ and are spanned by $|\vec\theta\rangle$ for $\vec\theta \in (\IR/2\pi\IZ)^d$ where 
\begin{equation}
    e^{i X_i} |\vec{\theta} \rangle \equiv e^{i\oint_{\gamma_i} \underline{a}} |\vec{\theta} \rangle = e^{i\theta_i} |\vec{\theta}\rangle\quad, \quad e^{i \vec\alpha\cdot \vec\CQ}|\vec\theta\rangle=|\vec\theta +2\pi\vec\alpha\rangle~,
\end{equation}
where $\vec\alpha\in (\IR/\IZ)^d$. 
Here we can then interpret the eigenvalues $e^{i\theta_j}$ as describing the holonomy of flat $U(1)$ connection on $T^d$. 

The other basis (Neumann) diagonalizes the $\CQ_i$ operators and is spanned by $|q_i\rangle$ for $\vec{q} \in 2\pi \mathbb{Z}^d$ where 
\begin{equation}\label{CCstates}
    \CQ_i |\vec{q}\rangle = q_i |\vec{q}\rangle\quad, \quad e^{i \vec{k} \cdot \vec{X}}|\vec{q}\rangle  = |\vec{q} -2\pi \vec{k} \rangle ~.
\end{equation} 
The two basis obey the standard orthonormality conditions:
\eq{\label{innerproductfromstates}
\langle \vec{q}|\vec{k}\rangle=\prod_i\delta_{q_i,k_i}\quad, \quad 
\langle \vec{\theta}|\vec{\phi}\rangle = \delta^d(\vec{\theta} - \vec{\phi}) ~.
}
Because these bases offer a resolution of the identity, the partition function will satisfy cutting and gluing conditions, as we would expect from a Lagrangian QFT, one the TQFT axioms set out in  \cite{Atiyah:1989vu}. 
However, we would like to point out that our SymTFT violates one axiom which requires the Hilbert space associated to a compact manifold $\Sigma_g$ to be finite dimensional. This is standard for example in $2d$ fusion category symmetries where the SymTFT is captured by its  Drinfeld center, which automatically satisfies the finiteness axiom and forms a modular tensor category (MTC). Therefore, one should be cautious in applying results there that follow from these axioms.

In our case, however, not only are the spectrum of the bulk line operators not finite (which is required by the definition of a MTC), but also they are parameterized by a continuous variable valued in $\mathbb{R}/\mathbb{Z}$ (together with an integer $\mathbb{Z}$). We will leave the proper description of the mathematical structure together with understanding how to generalize the known results on TQFTs satisfying the Atiyah axioms to future study.

These two bases are related by the Fourier series transform
\eq{
|\vec{\theta} \rangle = \frac{1}{(2\pi)^{d/2}} \sum_{\vec{q} \in 2\pi \mathbb{Z}^d} e^{\itwopi \vec{q} \cdot \vec{\theta}} |\vec{q}\rangle
}
Note that one can see that in both the Dirichlet and Neumann bases, that  the integer $\tildeh$-Wilson surfaces, here written as $e^{ i n_j\CQ_j}$, acts as the identity operator.

Quantizing the theory on a generic manifold $X_d$ will lead to a basis of states labeled by gauge inequivalent flat $U(1)$ connection $\underline{A}$'s on $X_d$ which diagonalize the Wilson lines operators
\begin{equation}
    e^{i\oint_\gamma \underline{a}} |\underline{A}\rangle = e^{i\oint_\gamma \underline{A}}|\underline{A}\rangle. 
\end{equation}
Acting with the Wilson surface operator $e^{i\alpha \oint_\Gamma \underline{\tildeh}}$ on the other hand will shift the background field by introducing an additional non-trivial holonomy $e^{2\pi i \alpha \,\#(\gamma\cap\Gamma)}$ along the $1$-cycle $\gamma$ which has non-trivial intersection with $\Gamma$.

\subsection{Gapped Boundaries of the SymTFT and Coupling to a QFT}\label{sec:U1gappedbdy}

We can also derive the allowed boundary conditions/states for the quiche from the Lagrangian perspective. This will be beneficial for studying the case where the SymTFT has additional couplings which arise for example in the case of QFTs with anomalous $U(1)$ global symmetries. 

The consistent boundary conditions of the theory are given by the (gauge invariant) sub-spaces of field space for which the boundary contribution to the action vanishes. For the $U(1)$ SymTFT, the variation of the action leads to boundary term
\begin{equation}\label{boundaryvarS}
    \delta S \big|_{bnd} = - \frac{i}{2\pi} \int_{X_d} \delta a \wedge \tildeh ~.
\end{equation}
In addition, the construction of the SymTFT requires quotienting by the gauge transformations that are non-trivial on the boundary due to the state-boundary correspondence.  Thus, we require that the gauge transformation of the action also vanishes:

\eq{\label{Sgaugbnd}
\delta_{gauge}S \big|_{bnd} = \frac{i}{2\pi} \int_{X_d} da \wedge \widetilde{\lambda} \quad, \quad \delta \tildeh=d\widetilde\lambda~.
}
The topological boundary conditions are therefore given by either 1.) fixing $a\big{|}_{bnd}=A$ to be a flat gauge field while $\tildeh$ fluctuates, or 2.) by fixing $\tildeh\big{|}_{bnd}=0$ while allowing $a$ to fluctuate\footnote{We can additionally fix $\tildeh\big{|}_{bnd}\neq 0$ by adding a boundary term as we will discuss later in this section.}. 
The first of these is the Dirichlet boundary condition while the second is the Neumann boundary condition.

Here, we focus on the following two type of topological boundary conditions:
\begin{enumerate}
    \item The path integral sums over $a$ such that $a\big|_{bnd} = A$ up to gauge transformation where $A$ is a flat connection, as well as $\tildeh$ such that $\tildeh\big|_{bnd}$ is flat and has integer holonomy;
    \item The path integral sums over $\tildeh$ such that $\tildeh\big|_{bnd} = 0$ up to gauge transformations and all flat connections $a\big{|}_{bnd}$. 
\end{enumerate}
The first of these is the Dirichlet boundary condition while the second is the Neumann boundary condition.

Let us focus on the Dirichlet boundary condition. {Here, the fact that we only fix the boundary condition}  $a\big|_{bnd} = A$ up to gauge transformation {is due to the fact that we require gauge invariant boundary conditions and matches} the result from the canonical quantization. The corresponding states, which we denote as $|A\rangle$, are labelled by gauge inequivalent flat $U(1)$ connections $A\in H^1(X_d,\mathbb{R}/\mathbb{Z})$ on $X_d$:
\begin{equation}
    |A\rangle \in \mathcal{H}_{X_d} \quad, \quad A \in H^1(X_d,\mathbb{R}/\mathbb{Z}) ~.
\end{equation}
{Thus, in the path integral we are only fixing $a\big{|}_{bnd}$ up to gauge transformations which  means that} the boundary variation of $a$ is given by a gauge transformation $\delta a \big|_{bnd}=d\varphi$ and the vanishing of the boundary variation of the action 
\eq{
\delta S\big{|}_{bnd}=-\itwopi\int_{X_d} \delta a\wedge \tildeh=-\itwopi\int_{X_d} d\varphi\wedge \tildeh~,
}
requires $d\tildeh\big|_{bnd} = 0$ and $\oint_\Gamma\tildeh\in 2\pi \IZ$ for $\Gamma \in C_{d-1}(X_d,\mathbb{Z})$. As a result, the $\tildeh$ Wilson surface operators are topological on the boundary and the integer $\tildeh$ Wilson surface operators $e^{in \oint_\Gamma \tildeh}$ {act as  trivial operators. }

We would like the derive the inner product of the Dirichlet boundary conditions from the path integral. 
Let us compute the partition function on $Y_{d+1}=X_d\times[0,1]$ and fix the boundary conditions $a\big{|}_{t=0,1}=A_{L,R}$ up to gauge equivalence. 

With these boundary conditions, we can rewrite the action as 
\eq{
S=\itwopi \int_Y a\wedge d\tilde{h}+\itwopi \int_{X_d} (A_L\wedge \tildeh_L-A_R\wedge \tildeh_R)~,
}
where $\tildeh_{L/R}:=\tildeh\big{|}_{t=0,1}$ respectively which are flat $\mathbb{R}$ connections with $2\pi \IZ$ holonomies.

Now we would like to compute the partition function with these boundary conditions: $\langle A_L|A_R\rangle$. Let us integrate out the bulk gauge field $a$. This imposes a constraint on $\tildeh$:
\eq{
\langle A_L|A_R\rangle=\int [d\tildeh]~\delta(d\tildeh)\,e^{\frac{i}{2\pi} \int_{X_d} (A_L\wedge\tildeh_L-A_R\wedge \tildeh_R)}~.
}
This delta function localizes the path integral to the constrained space of $\IR$-valued gauge fields $\tildeh$ which satisfy:
\eq{
\underline{\dot\tildeh}-\underline{d}\tildeh_t=0\quad , \quad \underline{d\tildeh}=0~. 
}
The first of these imposes that $\tildeh_L=\tildeh_R+\underline{d}\widetilde{\lambda}$ where $\widetilde{\lambda} = - \int_0^1 \, dt \, \tildeh_t$ and the second imposes $\tildeh_L\in Z^{d-1}(X_d;2\pi\IZ)$. The path integral then localizes:
\begin{equation}
\begin{aligned}
    \langle A_L | A_R\rangle &= \int [d \tildeh_L] ~e^{\frac{i}{2\pi} \int_{X_d} (A_L - A_R) \wedge \tildeh_L + A_L \wedge d\widetilde{\lambda}} \\
    &= \int [d\lambda]~ e^{\frac{i}{2\pi} \int_{X_d} (A_L - A_R) \wedge d\lambda} = \delta([A_L-A_R]) ~,
\end{aligned}
\end{equation}
where in the second step we replace the integral over flat connections with $2\pi\mathbb{Z}$ holonomy $\tildeh_L$ with the integral over $U(1)$ $(d-2)$-form connection $\lambda$. Notice that this is valid (up to a normalization factor) because $A_L - A_R$ is a flat connection, therefore only the flux sum $d\lambda$ would contribute non-trivially and reproduce the holonomy sum of $\tildeh_L$.  Here, we see that the inner product enforces that $A_L-A_R$ is the trivial cohomology class {which we denote as $\delta([A_L - A_R])$}, reproducing the inner product from canonical quantization in \eqref{innerproductfromstates}.

\bigskip

Now we would like to discuss the Neumann boundary condition. From the previous discussion, we expect this class of boundary condition is labelled by gauge inequivalent flat $\mathbb{R}$ connections $\tildeh\big|_{bnd}$ with $2\pi \mathbb{Z}$ {holonomies, {or equivalently, $(d-1)$-form $\mathbb{Z}$ gauge fields.} This holonomy can be} conveniently represented as the {flux part of the} field strength $dB$ of a boundary $U(1)$ $(d-1)$-form gauge field $B$. {To realize such a  boundary conditions, we} add the additional gauge invariant boundary term 
\eq{\label{eq:Svarbndm}
S_{bnd}=\itwopi \int a\wedge dB~.
}
 This modifies the boundary variation of the action 
\eq{
\delta S\big{|}_{bnd}=-\itwopi\int \delta a\wedge (\tildeh-dB)~, 
}
so that the Neumann boundary conditions are given by fixing 
\eq{
\tildeh\big{|}_{bnd}=dB~  \text{up to $\mathbb{R}$ gauge transformation}
}
while allowing $a$ to be {a general} flat $U(1)$ connection on the boundary, which guarantees the vanishing of \eqref{eq:Svarbndm} and the gauge variation \eqref{Sgaugbnd}.  Because of this, we can naturally identify the Neumann boundary condition
\eq{
|N_{dB}\rangle=  \int_{\CA_0/\CG}[dA_1]~e^{\itwopi \int_{X_d}A_1\wedge dB_{d-2}}~|A_1\rangle~,
}
where $\CA_0/\CG$ is the space of flat $U(1)$ connections modulo gauge transformations  
where $B_{d-2}$ is a $(d-2)$-form $U(1)$ gauge field.\footnote{{Here we will implicitly normalize our path integral by the (regulated) volume of $\CA_0/\CG=H^1(X_d;U(1))$, but will suppress the normalization factor for convenience. }
} {Note that the phase labeled by $dB_{d-2}$ is similar to the case of the standard (i.e. dynamical) gauging of the $U(1)$ symmetry except that for flat gauge fields the Neumann boundary condition only depends on the flux $dB_{d-2}$ instead of the full $(d-2)$-form gauge field $B_{d-2}$. }

This decomposition allows us to compute the inner product of the Neumann boundary conditions:
\eq{
\langle N_{dB}|N_{dB^\prime}\rangle&= \int_{\CA_0/\CG} [dA][dA']~\langle A|A'\rangle~e^{-\itwopi \int \left(A\wedge dB-A'\wedge dB'\right)}\\
&=\int_{\CA_0/\CG} [dA]~e^{\itwopi \int A\wedge (dB-dB')}
=\delta_{[dB],[dB']}~. 
}
Notice that because the $A$ integral is only taken over the space of flat $U(1)$ connections, it only sets equals the flux of the $B$ and $B'$ up to $\IR$ gauge transformations, which we denote as $\delta_{[dB],[dB']}$. {Note that these fluxes, and therefore the Neumann states themselves,}  are classified by a set of integers and have an inner product of the form of a Kronecker delta function as we found in the canonical quantization.

\bigskip

Now, let us describe the operator content of the SymTFT in the presence of the two boundary conditions. 
With the Dirichlet boundary condition $|A\rangle$, the $a$-Wilson lines are diagonalized as in the case of canonical quantization. Due to the linking of the $a$-Wilson lines and $\tildeh$-Wilson surfaces, we see that the $\tildeh$-Wilson surface acts non-trivially on the boundary state:
\eq{
e^{ i \alpha \oint_\Gamma \tildeh}|A\rangle=|A'\rangle\quad, \quad A-A'=2\pi \alpha \,\delta(\Gamma)~.
}
Additionally, as in the case of the $\IZ_N$ SymTFT, the Wilson lines can end terminate on the boundary to construct defect Hilbert spaces.

In the Neumann boundary condition, the $\tilde{h}$-lines are diagonalized by the boundary {state}. Due to the action of the $\tildeh$-surfaces on the Dirichlet states, we see that:
\eq{
e^{i \alpha \oint\tildeh}|N_B\rangle=e^{i \alpha \oint dB}|N_B\rangle~. 
}
The fact that $\oint \tildeh_d\in 2\pi\IZ$ is also reflected in the canonical quantization computation from the previous section as in \eqref{CCstates}. Similarly, the action of the $a$-Wilson line shifts:
\eq{
e^{i n \oint_\gamma a}|N_B\rangle\longmapsto |N_{B'}\rangle\quad, \quad dB'-dB=2\pi n\, \delta(\gamma)~.
}

\begin{figure}
    \centering
    \includegraphics[scale = 0.7]{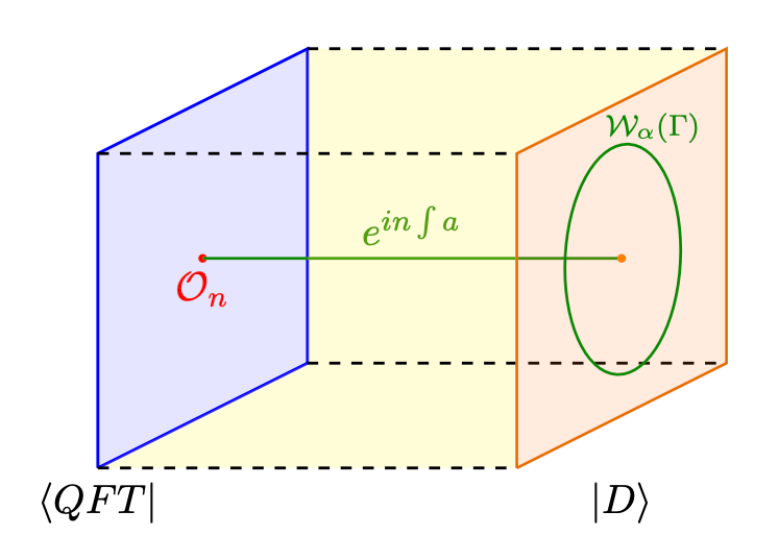}
    \caption{An operator $\mathcal{O}_n$ with charge $n$ under the $U(1)^{(0)}$ symmetry in the SymTFT is captured by a Wilson line stretching between $\mathcal{O}_n$ on the QFT boundary (blue) and a point on the Dirichlet boundary (orange). The Wilson surface $\CW_\alpha$ becomes the $U(1)$ symmetry operator.}
    \label{fig:Dir_new_1}
\end{figure}

\bigskip 

Now we are ready to describe how to couple the SymTFT to QFT. Let's consider a $d$-dim QFT $\mathcal{T}$ on $X_d$ with $U(1)^{(0)}$ global symmetry. In the SymTFT on $X_d \times [0,1]_t$ the QFT lives at $t=0$ while the quiche boundary lives at $t=1$. The QFT boundary is naturally described by a state
\begin{equation}
    \langle QFT| = \int_{H^1(X_d,\mathbb{R}/\mathbb{Z})} [dA] ~\mathcal{Z}_{\mathcal{T}}[A] ~\langle A| ~,
\end{equation}
where $Z_{\CT}[A]$ is the partition function of the theory $\mathcal{T}$ coupled to flat background $U(1)$ connection $A'$ on $X_d$.

Pairing the QFT state with a Dirichlet boundary state $|A\rangle$ on the quiche boundary
effectively leads to the inner product of the two boundary states $\langle {\rm QFT}| A \rangle $. Using the orthogonality of the Dirichlet states, we recover the partition function of the theory $\mathcal{T}$ coupled to the flat $U(1)$ connection $A$:
\begin{equation}
    \langle {\rm QFT}|A\rangle = Z_{\mathcal{T}}[A] ~.
\end{equation}

With the Dirichlet pairing, a local operator $\mathcal{O}_n$ in the QFT with charge $n$ under the $U(1)^{(0)}$ symmetry is captured a Wilson line that stretches across the slab so that the quiche boundary state is an element of the defect Hilbert space as shown in the Figure \ref{fig:Dir_new_1}. Here the action of a  $U(1)^{(0)}$ symmetry operator on $\mathcal{O}_n$ is captured by encircling the end point of the Wilson line on the Dirichlet boundary with the associated operator $\CW_\alpha(\Gamma)$.

Generically, in a QFT with $U(1)^{(0)}$ symmetry there are codimension-$2$ (non-topological) surface operators $\mathcal{S}_\alpha$ bounded by the corresponding $U(1)$ symmetry operator.\footnote{These operators will become the more familiar Gukov-Witten surface operators (which are also sometimes known as Aharanov-Bohm strings) in the phase where we gauge the $U(1)^{(0)}$ symmetry \cite{Gukov:2006jk,Gukov:2008sn}.} As a result, around the these operators, the background gauge field has holonomy $e^{i \alpha}$ where $\alpha \in U(1)$. In the sandwich picture, these $\CS_\alpha$ in the QFT boundary are constructed from a $\CW_\alpha(\Gamma)$ surface terminating on the QFT boundary. However, since  $\CW_\alpha(\Gamma)$ can not end on the Dirichlet boundary, the $\CW_\alpha$ operator must extend along the Dirichlet boundary when it reaches the end of the interval as shown in Figure \ref{fig:Dir_new_2}. After shrinking the sandwich, the tail of $\CW_\alpha$ on the Dirichlet boundary condition naturally becomes the $U(1)^{(0)}$ symmetry operator that bounds $\CS_\alpha$ in the QFT. 

\begin{figure}
    \centering
    \includegraphics[scale = 0.6]{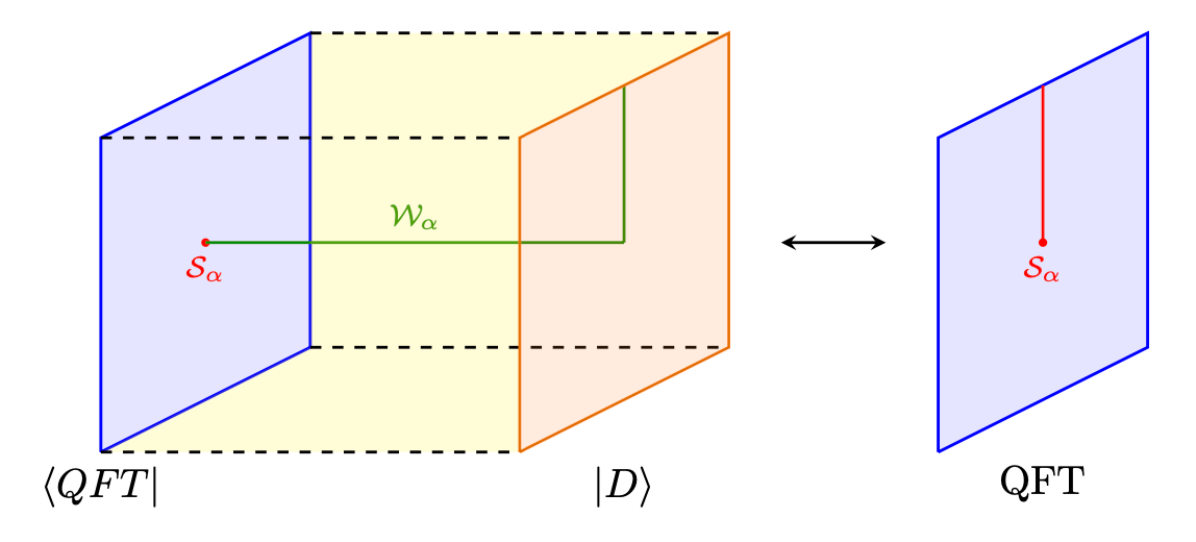}
    \caption{A non-local codimension-$2$ surface operator $\mathcal{S}_\alpha$ bounding the open $U(1)^{(0)}$ symmetry operator in the QFT is described by the Wilson surface $\CW_\alpha$ terminates on the operator $\mathcal{S}_\alpha$.}
    \label{fig:Dir_new_2}
\end{figure}
Similarly, we can couple the QFT state with the Neumann boundary condition:
\eq{
\langle {\rm QFT}|N_B\rangle=\int_{\CA_0/\CG}[dA]\, Z_\CT[A]\, e^{ \itwopi\oint A\wedge dB}~. 
}
This is the partition function where we have performed a ``flat gauging" of the $U(1)$ symmetry -- i.e. we have summed over only flat gauge $U(1)$ connections with a phase determined by a fixed choice of $\tildeh\big{|}_{bnd}\in H^{d-1}(X_d;\IZ)$ which we represent as 
the flux of the ``dual" $U(1)$ background gauge field $B$.  
While the above mathematical manipulation is  allowed, it is slightly unclear what the correct physical interpretation of such a gauging is. We will leave a discussion of such a gauging to future discussions. 

\subsection{Spontaneous Symmetry Breaking of  $U(1)$  Global Symmetry}
\label{sec:phases}

An important feature of the SymTFT is that it provides a tool which an be used to classify the possible IR phases of generic QFTs that realize a given symmetry structure. Due to its topological nature, the SymTFT is particularly well suited to classify the possible topological phases that can realize a certain categorical symmetry.\footnote{It is certainly an interesting question whether or not one can use the SymTFT to additionally classify the conformal phases that can realize a given symmetry. Classifying such conformal phases would correspond to classifying the conformal boundary conditions of the SymTFT. We will not classify these conformal boundary conditions here, but will give to a couple important examples. See \cite{Bhardwaj:2023bbf} for a discussion of conformal boundary conditions in the SymTFT for finite symmetries. }
These topological phases can be achieved by considering the possible (topological) states $\langle {\rm QFT}|$ and how they can be paired with the topological states of the quiche boundary \cite{Bhardwaj:2023idu,Bhardwaj:2023fca,Freed:2022qnc,Kaidi:2023maf,Kaidi:2022cpf,Kong:2020cie,Apruzzi:2021nmk,Bhardwaj:2023bbf,Apruzzi:2023uma,Bhardwaj:2023ayw,vanBeest:2022fss,Sun:2023xxv,Zhang:2023wlu,Cordova:2023bja,Antinucci:2023ezl,Cvetic:2023plv,Baume:2023kkf,Damia:2022bcd}. 

First, let us consider the case where $\langle QFT|=\langle N_B|$ with the quiche boundary condition $|A\rangle$. As in the case of the $\IZ_N^{(0)}$ SymTFT, since the two bases of boundary conditions are fourier transforms of each other, we see that the partition function: 
\eq{
\langle QFT|A\rangle=\langle N_B|A\rangle=\int_{\CA_0/\CG}[dA'] \langle A'|A\rangle\, e^{\itwopi \int_{X_d}A'\wedge dB}=e^{\itwopi\int_{X_d}A\wedge dB}~. 
}
encodes the trivial SPT phase for the $U(1)^{(0)}$ symmetry.\footnote{
Actually, the pairing $\langle N_{dB}|A\rangle$ is a non-trivial SPT of the $U(1)^{(0)}\times \IZ^{(d-2)}$ global symmetry where $\IZ^{(d-2)}$ is the ``dual" quantum symmetry described by the Neumann conditions. But it is the trivial SPT if we restrict to the $U(1)^{(0)}$ symmetry by fixing $dB_{d-2}=0$. 
}

Another phase of $U(1)$ global symmetries is the case when there is spontaneous symmetry breaking. In the case of a $\IZ_N$ global symmetry, the spontaneous symmetry breaking phase is described by the Dirichlet state. The reason is that the Dirichlet boundary conditions encapsulate the $\tildeh$-symmetry operators (which realize the domain walls) as well as charged operators (in the defect Hilbert spaces) which act as the order parameter.

For continuous symmetries, the arguments from $\IZ_N$ generalize straightforwardly. This implies that the QFT phase that realizes the spontaneous symmetry breaking should also be realized by a Dirichlet state, possibly dressed by a non-trivial phase. Indeed, we can see this by noting that the orthogonality relation for the Dirichlet-Dirichlet boundary conditions can be rewritten in the more suggestive way as 
\eq{
\langle A|A'\rangle=\delta([A-A'])=\int [d\varphi]~ \delta(A-A'-d\varphi)~, 
}
where $\varphi$ is a periodic scalar field corresponding to gauge transformation parameter of the $U(1)$ gauge field. Since $\varphi$ is a periodic scalar field $d\varphi\in H^1(X_d;\IZ)$ and we are imposing a Dirac delta function on the cohomology classes $[A_1'-A_1]\in H^1(M;U(1))$. 

Here we see that the inner product can be interpreted as the partition function over the field configuration space of a $U(1)$-valued Goldstone boson. This is suggestive that we should identify this phase with the spontaneous symmetry breaking (SSB) phase  of the $U(1)^{(0)}$ global symmetry without a kinetic term. 

In order to determine the correct $\langle{\rm QFT}|=\langle {\rm SSB}|$ to describe the spontaneous symmetry breaking, we would like to also match the partition function of the Goldstone mode by $\langle {\rm QFT}|A\rangle$. To incorporate the kinetic term, we consider the state
\begin{equation}\label{eq:SSBstate}
    \langle {\rm SSB}| = \int_{\mathcal{A}_0}[dA] ~e^{-\frac{1}{R^2} \int_{X_d} A\wedge *A} \langle A| =  \int_{\mathcal{A}_0/\CG}[dA]\int[d\varphi]~ e^{-\frac{1}{R^2}\int_{X_d}(d\varphi + A)\wedge *(d\varphi + A)} \langle A|~,
\end{equation}
and it's straightforward to check that
\begin{equation}
    \langle {\rm SSB}| A \rangle = \int [d\varphi] e^{-\frac{1}{R^2}\int_{X_d} (d\varphi + A)\wedge *(d\varphi + A)} ~,
\end{equation}
which reproduces the partition function of a Goldstone boson coupling to a flat $U(1)$ connection $A$.

As one can see, the $\langle {\rm SSB}|$ breaks the topological invariance, therefore does not represent a topological boundary condition of the SymTFT. This is not a surprise as the standard kinetic term for the Goldstone boson breaks topological invariance but preserves conformal symmetry, so $\langle {\rm SSB}|$ can be interpreted as a conformal boundary condition of the SymTFT. In a general QFT, the IR phase at finite energies will exhibit quantum corrections. Indeed, the standard $U(1)^{(0)}$ SSB phase will generically have higher order corrections as well as couplings to other dynamical sectors. These theories can still be consistently coupled to the $U(1)$ SymTFT due to the non-linearly realized $U(1)$ symmetry. They will have a different (i.e. non-conformal) $\langle {\rm QFT}|$ state which includes the higher order corrections. 

We conclude by mentioning that the state corresponding to the spontaneously symmetry breaking of $U(1)/\mathbb{Z}_N$ subgroup of $U(1)$ can be constructed similarly following \eqref{eq:SSBstate}
\begin{equation}
    \langle {\rm SSB}_N| = \int [dA] \int [d\varphi] e^{-\frac{1}{R^2}\int_{X_d}(d\varphi + N A) \wedge *(d\varphi + NA)} \langle A|~.
\end{equation}

\subsection{Global Form of Symmetry: $U(1)$ vs $U(1)/\IZ_N$}\label{sec:global_form}

Now let us discuss how the global form of the $U(1)$ global symmetry is realized. Here by the global form of the symmetry we mean fixing our global symmetry to be $U(1)$ (where the unit charge is $1$) v.s. $U(1)/\IZ_N$ (where the unit charge is $N$).

It is straightforward to write down the SymTFT for $U(1)/\mathbb{Z}_N$ symmetries, where we simply need to replace the $U(1)$ gauge field $a_1$ to be the $U(1)/\mathbb{Z}_N$ gauge field in the action \eqref{eq:SymTFT_action}. Notice that we can also start from a $U(1)$ SymTFT and gauge a $\mathbb{Z}_N^{(1)}$-form symmetry to get the $U(1)/\mathbb{Z}_N$ SymTFT.

To see this, one could rewrite the $U(1)^{(0)}$-SymTFT as a coupled theory between $(U(1)/\mathbb{Z}_N)^{(0)}$-SymTFT and a $\mathbb{Z}_N$-SymTFT:
\begin{equation}\label{eq:U(1)ZNaction}
    S = \frac{i}{2\pi} \int da_1 \wedge \tildeh_{d-1} + \frac{iN}{2\pi} \int dA_1 \wedge B_{d-1} - \frac{iN}{2\pi} \int da_1 \wedge B_{d-1}
\end{equation}
where $a_1$ is a $U(1)/\mathbb{Z}_N$ gauge field and $A_1, B_{d-1}$ are $U(1)$ gauge fields and $\tildeh_{d-1}$ is a $\mathbb{R}$ gauge fields. We want to emphasize here that because $a_1$ is a $U(1)/\mathbb{Z}_N$ gauge field, it is not possible to absorb $a_1$ into $A_1$ and the last term is indeed non-trivial.

Without the coupling term between $U(1)/\mathbb{Z}_N$-SymTFT and $\mathbb{Z}_N$-SymTFT, the spectrum of topological operators are given by
\begin{equation}
\begin{aligned}
    & e^{iqN \oint_{\gamma} a_1}, \quad e^{i\alpha \oint_{\Gamma} \tildeh_{d-1}}, \quad q \in \mathbb{Z}, \quad \alpha \in \left[0,1/N\right) ~, \\
    & e^{in \oint_{\gamma} A_1}, \quad e^{i n \oint_{\Gamma} B_{d-1}}, \quad n \in \mathbb{Z}_N, \quad n \in \mathbb{Z}_N ~.
\end{aligned}
\end{equation}
{When there is a coupling, the way the flux sums identify {operators} {is} modified.}  The flux sum of $\oint \frac{dB_{d-1}}{2\pi} \in \mathbb{Z}$, instead of identifying the charge $N$ $A_1$-Wilson line $e^{iN\oint A_1}$ with the trivial operator, now identifies that line with the minimal $a_1$-Wilson line (which is of charge $N$ as $a_1$ is a $U(1)/\mathbb{Z}_N$ connection)
\begin{equation}
    e^{i N \oint_\gamma A_1} \simeq e^{i N \oint_\gamma a_1} ~.
\end{equation}
Because of this, we can match the $U(1)$ line operator spectrum:
\eq{
W_n \simeq e^{i q \oint a_1}\times e^{i p N \oint A_1}\quad, \quad n=q+Np \quad \text{where} \quad q \in \{0,\cdots,N-1\}~.
}
Similarly, the flux sum of $\oint \frac{da_1}{2\pi} \in \frac{1}{N}\mathbb{Z}$ identifies 
\begin{equation}
    e^{\frac{i}{N}\oint \tildeh_{d-1}} \simeq e^{i \oint B_{d-1}}
\end{equation}
which extends the $U(1)/\mathbb{Z}_N$ to $U(1)$ and we can match the symmetry operators as
\begin{equation}
    \CW_\alpha \simeq e^{i\alpha' \oint \tildeh_{d-1}} e^{iq \oint B_{d-1}}\quad , \quad \alpha = \alpha' + \frac{q}{N} ~,
\end{equation}
where $\alpha' \in \left[0,1/N\right)$ and $ q \in \{0,\cdots,N-1\}$. 
It is straightforward to check that these identifications preserve the desired braiding relation to describe a $U(1)$-SymTFT. 

To get the $U(1)/\mathbb{Z}$-SymTFT itself, we only need to gauge the $\mathbb{Z}_N^{(1)}$-symmetry generated by the surface operator $e^{i \oint B_{d-1}}$. This will project out all the $A_1$-Wilson lines, thereby effectively setting $B_{d-1} = A_1 = 0$ in the action \eqref{eq:U(1)ZNaction} and resulting in the  $(U(1)/\mathbb{Z}_N)^{(0)}$-SymTFT.

The form of $U(1)^{(0)}$-SymTFT \eqref{eq:U(1)ZNaction} also allows us to naturally describe some other boundary conditions one can get with $U(1)^{(0)}$-SymTFT. For instance, starting with $U(1)^{(0)}$ global symmetry and gauging a $\mathbb{Z}_N^{(0)}$ subgroup will lead to the symmetry $(U(1)/\mathbb{Z}_N)^{(0)} \times \mathbb{Z}_N^{(d-2)}$-form symmetries with a mixed 't Hooft anomaly. The new symmetry is described by the same $U(1)^{(0)}$-SymTFT, and the discrete gauging is simply realized by picking different boundary states\footnote{Note that the general boundary condition (including this discrete gauging) can come with an additional topological term for the $\IZ_N^{(0)}$ gauge field and the $U(1)$ gauge field. We will not consider these additional terms in this paper. }
\eq{\label{NeumannU1ZN}
\big{|}A_1;B_{d-1}\big{\rangle}_{U(1)/\IZ_N}:=\frac{1}{\sqrt{|H^1(X_d,\mathbb{Z}_N)|}} \sum_{A_1' \in H^1(X_d,\frac{2\pi}{N}\mathbb{Z}_N)} e^{\frac{iN}{2\pi}\int_{X_d} A_1' \cup B_{d-1}} 
    \big{|}\hat{A}_1 + A_{1}'\big{\rangle}_{U(1)}~.
}
where $A_1$ is a $U(1)/\IZ_N$ gauge field with a choice of $U(1)$ lift $\hat{A}_1$, and the field $B_{d-1}$ is the background gauge field for the dual $\IZ_N^{(d-2)}$ symmetry. On the other hand, the description of the $U(1)$ SymTFT in \eqref{eq:U(1)ZNaction} is more natural for simultaneously realizing the $(U(1)/\IZ_N)^{(0)}$ and  dual $\IZ_N^{(d-2)}$  symmetry: the boundary condition realizing $(U(1)/\mathbb{Z}_N)^{(0)}\times \mathbb{Z}_N^{(d-2)}$ global symmetries is the one generated by condensing  the $a_1$-Wilson lines together with the $B_{d-1}$-Wilson surface operators and  the coupling term characterizes the mixed 't Hooft anomaly between $(U(1)/\mathbb{Z}_N)^{(0)}$ and $\mathbb{Z}_N^{(d-2)}$.

\subsection{Non-Flat Connections and Dynamical $U(1)$ Gauging}
\label{sec:dynamicalgauging}

In this subsection, we will describe how to realize non-flat connection in the SymTFT and also give a description of dynamical $U(1)$ gauging in the SymTFT. {We want to warn the readers that the construction here is quite different from the one in conventional SymTFT: the boundary states corresponding to the non-flat connections are not states in the SymTFT Hilbert space, but rather states in some defect Hilbert space. Subsequently, the sandwich description of the dynamical $U(1)$ gauging is achieved by coupling a $U(1)$ gauge theory to the SymTFT where the monopole (charged operators) and the symmetry operators of the dual $U(1)^{(d-3)}$ magnetic symmetry both live on the boundary, but can be pushed into the SymTFT bulk by a change of variables.}

We will first describe how we can realize Dirichlet boundary conditions with non-flat connection. As discussed previously, imposing the Dirichlet boundary condition $a_1|_{bdy} = A_1$ where $dA_1 \neq 0$ violates the gauge invariance due to a surface term under gauge transformation. On the other hand, it is possible to cancel this by adding a bulk term. 

To see this, we go back to the action in terms of decomposed fields on $X_d \times [0,1]_t$: 
\begin{equation}
    S = \frac{i}{2\pi} \int_{0}^{1} dt \int_{X_d} \dot{\underline{a}} \wedge \underline{\widetilde{h}} + a_t \underline{d}\underline{\tildeh} + \underline{da} \wedge \tildeh_t ~, 
\end{equation}
and consider adding an extra bulk term
\begin{equation}\label{eq:bulk_defect_term}
    \Delta S = - \frac{i}{2\pi} \int_0^1 dt \int_{X_d} \underline{dA} \wedge \tildeh_t ~,
\end{equation}
where $\underline{A}$ is a generic $U(1)$ connection along $X_d$-direction and does not depend on $t$. Adding $\Delta S$ does not affect the boundary term arising from the variation of the action, but will introduce non-trivial surface term under the gauge transformation to ensure gauge invariance when the boundary value $a_1$ is non-flat. To see this, notice that the gauge transformation of the decomposed fields are
\begin{equation}
\begin{aligned}
    & \underline{\tildeh} \rightarrow \underline{\tildeh} + \underline{d} \underline{\tildelambda} ~, \quad \tildeh_t \rightarrow \tildeh_t - \underline{d}\tildelambda_t + \dot{\underline{\tildelambda}} ~, \\
    & \underline{a} \rightarrow \underline{a} + \underline{d}\varphi~, \quad a_t \rightarrow a_t + \dot{\varphi} ~,
\end{aligned}
\end{equation}
where we decompose the gauge transformation parameter $\lambda$ for $\tildeh \rightarrow \tildeh + d\tildelambda$ as
\begin{equation}
    \tildelambda = dt\wedge \tildelambda_t + \underline{\tildelambda} ~.
\end{equation}
Then, we find, up to total derivative along $X_d$ direction,
\begin{equation}
\begin{aligned}
     \delta_{gauge} S& = \frac{i}{2\pi} \int_{0}^1 dt \int_{X_d} \frac{\partial}{\partial t}(\underline{da} \wedge \underline{\tildelambda}) ~, \\
     \delta_{gauge} (\Delta S)& = - \frac{i}{2\pi} \int_0^1 dt \int_{X_d} \frac{\partial}{\partial t} (\underline{dA} \wedge \underline{\tildelambda}) ~,
\end{aligned}
\end{equation}
so that $S + \Delta S$ is gauge invariant. Thus, one can realize a non-flat $U(1)$ connection as the Dirichlet boundary condition of $a_1$ on the boundary by adding the term \eqref{eq:bulk_defect_term}.

In the special case where $\frac{1}{2\pi}\underline{dA}$ is the Poincare dual of some $(d-2)$-cycle $\Sigma \in H_{d-2}(X_d,\mathbb{Z})$, the bulk term
\begin{equation}
    e^{-\Delta S} = e^{\frac{i}{2\pi} \int_0^1 dt \int_{X_d} \underline{dA} \wedge \tildeh_t} = e^{i \int_0^1 dt \int_{\Sigma} \tildeh_t} = e^{i \int_{\Sigma \times [0,1]_t} \tildeh}
\end{equation}
is nothing but an integer $\tildeh$ Wilson surface extending along the $t$-direction -- more generally it is a smeared $\tildeh$-Wilson surface given by \eqref{eq:bulk_defect_term}. This means we should interpret the boundary states representing the $U(1)$ connections whose field strength is $2\pi \delta_2(\Sigma)$ should be viewed as a state in the defect Hilbert space of the operator $e^{i \int_{\Sigma \times [0,1]_t} \tildeh}$. 
Based on this, we then view a generic state representing a non-flat connection $A_{nf}$ as a state $|A_{nf}\rangle\rangle$ in the defect Hilbert space $\mathcal{H}_{F}$ where $F$ is the field strength associated to the connection $A_{nf}$. Here use the notation $|A\rangle\rangle$ to differentiate states in defect Hilbert spaces $\CH_F$ from the  Hilbert space of the theory.

Notice that if $|A_{nf}\rangle\rangle$ and $|A_{nf}'\rangle\rangle$ belong to the same defect Hilbert space $\mathcal{H}_F$, then  $A_{nf}' - A_{nf}$ is a flat connection, as the corresponding $\Delta S$ for the two connections are identical. Their inner product is then computed identically to before except that there is a non-zero base-point connection $A_{nf}$.

{Proceeding as before, we find that after imposing large gauge invariance on the boundaries, the inner product on a defect Hilbert space $\mathcal{H}_{F}$  is given by }
\begin{equation}
	\langle\langle  {A'}|  {A} \rangle\rangle = \int [d\lambda] \, \exp\left(\frac{i}{2\pi} \int_{X_d} (A' - A) \wedge d\lambda \right) =\delta([A-A^\prime])~,
\end{equation}
{where $|A\rangle\rangle,|A'\rangle\rangle\in \CH_F$ and }$\lambda$ is $(d-2)$-form $U(1)$ gauge field which serves as a Lagrangian multiplier to set $A' = A$ up to gauge transformation.

{
We can then define the extended Hilbert space $\widehat{\CH}=\bigoplus_F\CH_F$ as the formal sum over all defect Hilbert spaces. The inner product on each $\CH_F$ then lifts to $\widehat{\CH}$ as 
\eq{
\langle\langle A|A'\rangle\rangle=\int [d\lambda]~e^{\itwopi\int_{X_d}(A'-A)\wedge d\lambda}~=\delta([A-A'])~,
}
for $|A\rangle\rangle,|A'\rangle\rangle\in \widehat{\CH}$. 
The QFT state can be naturally extended to a state in $\widehat{\CH}$ by incorporating the non-flat connections as:}
\begin{equation}\label{eq:extendedQFT}
    \langle \langle {\rm QFT} | = \int_{\mathcal{A}/\mathcal{G}} [dA'] Z_{\mathcal{T}}[A'] e^{-\frac{1}{2g^2}\int_{X_d} dA' \wedge * dA'} \langle\langle A'| ~,
\end{equation}
where  $\mathcal{A}/\mathcal{G}$ is the space of all $U(1)$ connections modulo  gauge transformations. 

In a similar spirit, one can define an ``extended Neumann" state
\begin{equation}\label{eq:Neumannse}
    |N_B\rangle \rangle = \int_{\mathcal{A}/\mathcal{G}} [dA]~ e^{\frac{i}{2\pi} \int dA \wedge B} |A\rangle \rangle~,
\end{equation}
where $B_{d-2}$ is a background $(d-2)$-form $U(1)$ gauge field. {Notice that since the extended Neumann state sums over non-flat gauge fields $A$, the state $|N_B\rangle\rangle$ depends on the the full data of the $(d-2)$-form gauge field $B_{d-2}$. 
}

Evaluating the inner product then leads to the partition function of the theory $\mathcal{T}$ with $U(1)$ dynamically gauged 
\begin{equation}\label{eq:Neum_inner_product}
\begin{aligned}
    \langle \langle &{\rm QFT} | N_B \rangle \rangle = \int_{\mathcal{A}/\mathcal{G}} [dA'][dA] ~Z_{\mathcal{T}}[A'] ~e^{-\frac{1}{2g^2}\int_{X_d} dA' \wedge * dA' + \frac{i}{2\pi} \int_{X_d} dA \wedge B}~ \langle\langle A'|A\rangle\rangle \\
    &= \frac{1}{\mathcal{N}}\int_{\mathcal{A}/\mathcal{G}} [dA'][dA] ~Z_{\mathcal{T}}[A'] e^{-\frac{1}{2g^2}\int_{X_d} dA' \wedge * dA' + \frac{i}{2\pi} \int_{X_d} dA \wedge B} e^{\frac{i}{2\pi}\int_{X_d}(A'-A)\wedge d\lambda} \\
    &= \int_{\mathcal{A}/\mathcal{G}} [dA] ~Z_{\mathcal{T}}[A]~ e^{-\frac{1}{2g^2}\int_{X_d}dA\wedge \ast dA + \frac{i}{2\pi}\int_{X_d} dA\wedge B} =Z_{\CT/U(1)}[B]~.
\end{aligned}
\end{equation}
This inner product has the bulk interpretation of summing over all insertions of integer $\tildeh$-Wilson surfaces with fixed end points on the QFT and quiche boundaries up to bulk topological deformations.

In order to study the operators in this theory, we can rewrite the path integral as
\begin{equation}\label{eq:Neum_ip_rw}
\begin{aligned}
    \langle \langle &{\rm QFT} | N_B \rangle \rangle = \int [dA'][dA][da][d\tildeh] Z_{\mathcal{T}}[A'] e^{-\frac{1}{2g^2}\int_{X_d} dA'\wedge * dA' + \frac{i}{2\pi} \int_{X_d} dA \wedge B + \frac{i}{2\pi} \int_{X_d \times [0,1]_t} (da - {dA}) \wedge \tildeh }
\end{aligned}
\end{equation}
where the path integral over $A,A'$ is taken over all $U(1)$ connections on $X_d$ and the path integral over $a$ is taken over all the $a$'s with the boundary condition $a|_{t=0} = A'$ and $a|_{t=1} = A$ up to a gauge transformation. {Notice on the domain where $A'-A$ is not flat, then the SymTFT part of the action is not gauge invariant and therefore the total contribution to the partition function on this domain will vanish identically. }

Here, we see that the bulk theory is modified by the appearance of an extra dynamical field $A$ along the $X_d$ direction, which leads to a TQFT like picture of the magnetic quantum $U(1)^{(d-3)}$ symmetry. Namely, the {charge $n$} codim-$3$ monopole operator placed on $\Sigma \subset X_d$ in the QFT can be lifted to a {(non-topological)} codim-$3$ monopole operator {for $a_1$}  supported on $\Gamma=\Sigma \times [0,1]_t$ in the bulk that lives at the end of a charge $n$ $\tildeh$-surface operator. The surface operator $e^{i\alpha \oint_{\sigma } dA'}$ wrapping  $\sigma \subset X_d$ then  lifts to the non-trivial operator  $e^{i\alpha \oint_{\sigma\times \{t_0\}} da}$ (due to its linking with the bulk monopole operator) and plays the role of the $U(1)^{(d-3)}$ symmetry operator. The operator $e^{ i \alpha \oint_{\sigma\times \{t_0\}} da}$ can then be pushed to the quiche boundary ($t_0\to 1$) where it becomes the operator $e^{i \alpha \oint_\sigma dA}$. Because of this, we can physically interpret the state $|N_B\rangle\rangle$ as introducing a free $U(1)$ gauge theory on the quiche boundary (described by $A$) which is then identified with the dynamical degrees of freedom of the $\CT/U(1)$ theory (described by $A'$) by computing the path integral over the bulk degrees of freedom (described by $a,\tildeh$). 

Although the boundary state $|N_B\rangle\rangle$ is not a standard boundary condition in the SymTFT since it is a formal sum over boundary conditions with bulk operators inserted, this construction is still useful for understanding $U(1)$ global symmetries in QFTs as we will demonstrate in the next section. 
We will leave the search for a SymTFT which realizes all $U(1)$ connections as genuine states in the Hilbert space to future study. 

\section{Applications}
\label{sec:applications}

In this section we discuss several applications and extensions of our construction of the Symmetry TQFT for $U(1)$ global symmetries.

\subsection{$U(1)^{(0)}$ `t Hooft Anomalies}
\label{sec:anomaly}

First we would like to discuss how anomalies of $U(1)$ global symmetries are incorporated into the $U(1)$-SymTFT. Here we will focus on the cubic anomaly of a single $U(1)^{(0)}$ in a $4d$ QFT which is given by the $5d$ anomaly SPT:\footnote{
The $5d$ anomaly SPT phase is the Chern-Simons term whose variation is a boundary term which describes the variation of the partition function. In terms of the descent formalism, the derivative of the $5d$ SPT action is the ``anomaly polynomial" which is  an integral-quantized characteristic class. 
}
\eq{
\CA = \frac{i\kappa}{24\pi^2} \int_{Y_5} a_1 \wedge da_1 \wedge da_1~.}
In the SymTFT such an anomaly is encoded by adding an analogous Chern-Simons term so that the total bulk action is
\eq{\label{eq:anomalous_U1} S = \frac{i}{2\pi} \int_{Y_5} da_1\wedge \widetilde{h}_3 + \frac{i\kappa}{24\pi^2} \int_{Y_5} a_1 \wedge da_1\wedge da_1~.}
This additional term has several effects. First, let's consider the operators in the bulk. The $5d$ CS term modifies the equation of motion of $a_1$ to be
\eq{\label{BulkU1cube}
d\widetilde{h}_3 = \frac{\kappa}{8\pi}da_1\wedge da_1~. }
Generically, this implies that the $\tildeh$-surface operator $e^{i\alpha \oint_{\Gamma} \tildeh_{3}}$ is no longer topological in the bulk. Furthermore, similar to the case of $\mathbb{Z}_N$-SymTFT discussed in Section \ref{sec:ZN_anomaly}, the surface operator $e^{i\alpha \oint_{\Gamma} \tildeh_{3}}$ in the bulk now has non-trivial self triple intersection. This implies that we can not consistently construct the Neumann boundary condition by the condensation of these operators on the boundary  \cite{Kaidi:2023maf,Cordova:2023bja,Zhang:2023wlu} and therefore we can not realize a trivially gapped phase via pairing with the Dirichlet boundary with the Neumann boundary. 

To demonstrate this as well as  other effects of the $5d$ CS term, we study the SymTFT placed on a manifold with boundary. The surface terms on the boundary from the variation of the action and the gauge variation are given by
\begin{equation}\label{eq:U1anomalyvar}
\begin{aligned}
    \delta S\Big|_{bnd} &= \frac{i}{2\pi} \int_{X_4} \delta {a_1} \wedge \left({\tildeh}_3 + \frac{\kappa}{6\pi} {a}_1 \wedge {da_1}\right)~, \\
    \delta_{\rm gauge}S\Big|_{bnd} &=\itwopi \int_{X_4} da_1\wedge \tildelambda_2+\frac{i\kappa}{24\pi^2}\int_{X_4} \varphi\, da_1\wedge da_1~, 
\end{aligned}
\end{equation}
where the gauge transformation is given by $a_1 \rightarrow a_1 + d\varphi, \, \tildeh_3 \rightarrow \tildeh_3 + d\tildelambda_2$. Notice that the anomaly term leads to a boundary contribution to the gauge variation of the action.

On the other hand, if the boundary value $a_1\big|_{bnd}$ is flat, then the extra contribution due to the $5d$ CS term vanishes. Therefore, it is straightforward to define a Dirichlet boundary condition realizing flat $U(1)$ background gauge field following the previous construction in Section \ref{sec:U1gappedbdy}. Furthermore, the vanishing of $da_1$ also guarantees that the $e^{i\alpha \oint_{\Gamma} \tildeh}$ on the boundary is topological and realizing the $U(1)$ global symmetries. 

{One way to see that} this anomaly obstructs the Neumann boundary condition for $a_1$ is the following. From \eqref{eq:U1anomalyvar}, we see that the Neumann boundary conditions, {which are  described by the solutions to}
\eq{\tildeh_3+\frac{\kappa}{6\pi}a_1\wedge da_1\Big{|}_{bnd}=0~,
}
are not compatible with the bulk equation of motion \eqref{BulkU1cube}. Thus we recover the well-known fact that the anomaly of $U(1)$ global symmetry prevents the existence of trivially gapped phase realizing anomalous $U(1)$ symmetry \cite{Coleman:1982yg}.

Now, we want to demonstrate that the SymTFT \eqref{eq:anomalous_U1}, when turning on non-flat connections on the boundary, produce the anomalous phase familiar in the $4d$ QFT. In this case, we must carefully define the path integral. With the bulk defect term  \eqref{eq:bulk_defect_term} {to source a non-flat boundary gauge field}, the variation of the action is modified to 
\begin{equation}
\begin{aligned}
    \delta (S+\Delta S)\Big|_{bnd} &= \frac{i}{2\pi} \int_{X_4} \delta {a_1} \wedge \left({\tildeh}_3 + \frac{\kappa}{6\pi} {a}_1 \wedge {da_1}\right)~, \\
    \delta_{\rm gauge}(S+\Delta S)\Big|_{bnd} &=\itwopi \int_{X_4} (da_1 - dA_1)\wedge \tildelambda_2+\frac{i\kappa}{24\pi^2}\int_{X_4} \varphi\, da_1\wedge da_1~. 
\end{aligned}
\end{equation}
Naively, one may want to define the path integral as summing over the bulk $a_1$ gauge field such that $a_1\big|_{bnd} = A_1$ up to gauge transformations. However, this leads to non-vanishing surface term under gauge variation as
\begin{equation}
    \delta_{gauge}(S+\Delta S)\Big|_{bnd} = \frac{i\kappa}{24\pi^2} \int_{X_4} \varphi \, dA_1 \wedge dA_1 \neq 0 ~,
\end{equation}
for generic $A_1$ and $\varphi$.

{To construct the gauge invariant quiche state,} one can start with the path integral where one sums over all bulk gauge field $a_1$ such that $a_1\big|_{bnd} = A_1$, and sum over all the 
gauge transformations $A_1\mapsto A_1+d\varphi$.

Notice that a {boundary gauge transformation can be described by a bulk} gauge transformation $a_1 \rightarrow a_1 + d\varphi$ {where $d\varphi\big{|}_{bnd}\neq0$. Such a gauge transformation} will shift $a_1|_{bnd}$ to $a_1|_{bnd} + d\varphi|_{bnd}$, { and} therefore relates different strict Dirichlet boundary conditions ({where we do not sum over boundary gauge transformations in the path integral and fix the boundary value of $a_1$ exactly) which we denote $|A\rangle\rangle_0$. Note that these are not physical states in the Hilbert space as we have not yet imposed gauge invariance.} 
{This allows us to compute that the strict Dirichlet states transform with a phase under a boundary gauge transformation:}
\eq{
|A\rangle\rangle_0&=\int\underset{a|_{bnd}=A}{[da\,d\tildeh]} e^{-S[a,\tildeh]}=\int\underset{a|_{bnd}=A}{[da\,d\tildeh]} e^{-S[a+d\varphi-d\varphi,\tildeh]}
=\int\underset{a'|_{bnd}=A+d\varphi}{[da'\,d\tildeh]} e^{-S[a'-d\varphi,\tildeh]}\\&
=e^{\frac{i\kappa}{24\pi^2}\int_{X_4}\varphi dA_1 \wedge dA_1}\int\underset{a'|_{bnd}=A+d\varphi}{[da'\,d\tildeh]} e^{-S[a',\tildeh]}
\\
&=e^{\frac{i\kappa}{24\pi^2}\int_{X_4}\varphi dA_1 \wedge dA_1}|A+d\varphi\rangle\rangle_0~.
}

{In order to construct gauge invariant states, we must integrate over the gauge orbit of a  strict Dirichlet states with an additional anomalous phase:}
\eq{
|A\rangle\rangle=\int [d\varphi] ~e^{\frac{i \kappa}{24\pi^2}\int_{X_4}\varphi\, dA_1\wedge dA_1}|A+d\varphi\rangle\rangle_0~.
}
{In the setting of the SymTFT, this phase naturally arises from pairing the gauge-dependent state with a partition function that exhibits the same anomalous phase.} 
This does not affect the construction of the extended QFT state \eqref{eq:extendedQFT} {since }the theory $\mathcal{T}$ has the corresponding 't Hooft anomaly. Namely, the combination $\langle \langle A_1 | Z_{\mathcal{T}}[A_1]$ is gauge invariant provided that $Z_{\mathcal{T}}[A_1 + d\chi] = Z_{\mathcal{T}}[A_1] e^{-\frac{i\kappa}{24\pi^2}\int_{X_4}\chi dA_1 \wedge dA_1}$, therefore the extended QFT state 
\begin{equation}
    \langle \langle {\rm QFT} | = \int_{\mathcal{A}} [dA'_1] ~Z_{\mathcal{T}}[A'_1] ~e^{-\frac{1}{2g^2}\int_{X_d} dA'_1 \wedge * dA'_1} \,_0\langle\langle A'_1| 
\end{equation}
remains well-defined. The extended {Dirichlet state for the quiche boundary can additionally be cured by dressing the state with a the bulk SPT phase as described in \cite{Freed:2022qnc}:}
\eq{
|A\rangle\rangle:=\int [d\varphi]~Z_{SPT}[A+d\varphi]~|A+d\varphi\rangle\rangle_0~.
}
Physically, this is analogous to the statement that the SymTFT is Witt equivalent to the anomaly SPT phase by condensing Wilson lines, for our case where the topological theory has an infinite-dimensional Hilbert space/set of Wilson lines.

{
The inner product between $\langle\langle QFT|$ and $|A\rangle\rangle$ then computes the gauge invariant combination of the partition function that is dressed by the $(d+1)$-dimensional SPT phase:}
\eq{
\langle \langle {\rm QFT}|A\rangle=Z_{\rm QFT}[A]\times Z_{SPT}[A]~.
}
The extended Neumann state \eqref{eq:Neumannse}, on the other hand, is ill-defined, consistent with the fact that one can not dynamically gauge the $U(1)$ symmetry when there is an anomaly. 

\subsection{Mixed $U(1)^{(0)}\times U(1)^{(0)}$ Anomaly and Non-Invertible $\IQ/\IZ$ Symmetry}

In a $4d$ theory with $U(1)^{(0)}_A\times U(1)^{(0)}_a$ global symmetry we can write down the SymTFT as 
\eq{
S_{U(1)\times U(1)}=\itwopi \int da_1\wedge \tildeh_3+ dA_1 \wedge \tildeH_3~. 
}
These symmetries admit a mixed anomalies: without loss of generality we will consider the case where the theories have a mixed $U(1)_a^2\times U(1)_A$ anomaly. This anomaly will prevent the existence quiche boundary conditions which realizes the symmetries as simultaneously gauged in the QFT. However, it does not prevent boundary conditions which realizes the $U(1)_a^{(0)}$ or $U(1)_A^{(0)}$ as gauged in the QFT. However, in the phase where $U(1)_a^{(0)}$ is gauged, the $U(1)_A^{(0)}$ will exhibit an ABJ anomaly.\footnote{Alternatively, in the phase where $U(1)_A$ is gauged, $U(1)_a$ participates in a 2-group \cite{Cordova:2018cvg,Benini:2018reh}. We will not discuss this scenario in this paper. }  As discussed in \cite{Cordova:2022ieu,Choi:2022jqy}, a $U(1)^{(0)}$ global symmetry in a $4d$ QFT with an ABJ anomaly of this type should be converted into a non-invertible $(\IQ/\IZ)^{(0)}$ global symmetry. Here we will demonstrate how the $U(1)^{(0)}\times U(1)^{(0)}$ SymTFT will capture this non-invertible symmetry with the appropriate choice of quiche boundary conditions.

In the SymTFT, the $U(1)_a^2\times U(1)_A$ anomaly can be accounted for by adding the term 
\eq{
\Delta S=ik\int A_1\wedge \frac{da_1\wedge da_1}{8\pi^2}~.
}
When we add this coupling, the allowed boundary variation is modified:
\eq{\label{QZboundvar}
\delta S\big{|}_{bnd}=\itwopi\int \delta a_1\wedge \left(\tildeh_3+\frac{k}{2\pi}A_1\wedge da_1\right)+\delta A_1\wedge \tildeH_3~. 
}
As in the case of the $U(1)$ self-anomaly, these boundary conditions do not allow  simultaneous Neumann boundary conditions for $a_1,A_1$. 

The anomaly also changes the bulk equations of motion for  $\tildeH_3,\tildeh_3$:
\eq{
d\widetilde{H}_3+\frac{k}{4\pi} da_1\wedge da_1 =0\quad, \quad d\widetilde{h}_3+\frac{k}{2\pi} dA_1\wedge da_1=0~. 
}
These are not compatible with the Neumann boundary conditions described by the boundary variation of the action in \eqref{QZboundvar}. We would like to comment that one can add a boundary term to the action which allows us to choose either the $\tildeh$ or $\tildeH$ equations of motion to be compatible with the corresponding Neumann boundary condition. However,  there does not exist a boundary term that makes both of them simultaneously compatible -- this is prevented by the term describing the anomaly. 

On a closed manifold without boundary the $\tildeh_3,\tildeH_3$ surfaces are topological due to the other equations of motion:
\eq{
\frac{da_1}{2\pi}=0\quad, \quad \frac{dA_1}{2\pi}=0~. 
}
However, in the presence of a boundary we can turn on $da_1,dA_1\neq 0$ in which case the $\tildeh_3,\tildeH_3$-surfaces may not be topological.

Here we will consider fixing the Dirichlet boundary condition for $A_1$ so that $A_1$ is a flat gauge field.
In the case with flat Dirichlet boundary conditions for $a_1$, the $\tildeH_3$ surface is topological. However, for generic 
boundary values of $a_1$ -- such as in a generic defect Hilbert space where $a_1$ is not flat --  the $\tildeH_3$ surfaces are not topological except for the surfaces of the form  $e^{\frac{in}{k}\oint (\tildeH_3+\frac{ik}{4\pi} a_1\wedge da_1)}$.  This is consistent with the fact that upon dynamically gauging the global $U(1)^{(0)}_a$ symmetry, the corresponding ABJ anomaly  will break the group-like symmetry $U(1)_A\mapsto \IZ_k$.   
 
However, as discussed in \cite{Cordova:2022ieu,Choi:2022jqy}, the ABJ anomaly for a $U(1)^{(0)}$ global symmetry transmutes the broken group-like symmetry into a non-invertible $\IQ/\IZ^{(0)}$ global symmetry. 

To realize this $\mathbb{Q}/\mathbb{Z}$ non-invertible symmetry in the $U(1)^2$ Symmetry TQFT, we can construct the topological operator associated to the $\tildeH$-surface by dressing the bare $\tildeH$-surface with a fractional quantum hall state 
\eq{
\CD_q[\Sigma]=\CA^{N,p}[\Sigma;a_1]\times e^{i q\oint_\Sigma \widetilde{H}_3}\quad, \quad k\,q=\frac{p}{N}~,
}
where $\CA^{N,p}[a_1]$ is the minimal $\IZ_N$ TQFT \cite{Hsin:2018vcg} which satisfies
\eq{
\delta_\Sigma \CA^{N,p}[\Sigma;a_1]=\CA^{N,p}[\Sigma;a_1]\times e^{-ik \int_{\delta \Sigma} \frac{a_1\wedge da_1}{4\pi}}~.
}
This composite operator $\CD_q[\Sigma]$ is topological as the non-topological nature of the $\tildeH_3$-Wilson surface and $\CA^{N,p}[\Sigma;a_1]$ cancel. However, due to the non-trivial structure of the product of the $\CA^{N,p}[\Sigma;a_1]$ operators \cite{Hsin:2018vcg}, the $\CD_q[\Sigma]$ will now generate a non-invertible symmetry structure \cite{Cordova:2022ieu,Choi:2022jqy}. 

This operator $\CD_q[\Sigma]$ is innately topological (i.e. independent of the boundary condition). However, when we take $a_1$ to have (flat) Dirichlet boundary conditions, the operator factorizes into the product of two topological operators -- one of which is the group-like $\tildeH_3$-Wilson surface. 

\section{Comments on Continuous Non-Abelian 0-form Symmetries}
\label{sec:nonabelian}

In this section we will propose a Symmetry TQFT for a non-abelian, continuous 0-form global symmetry. Our proposal is a simple extension of the $U(1)$ SymTFT where we interpret $\IR$ as the Lie algebra of $U(1)$.

Let us take $G$ to be a continuous non-abelian Lie group and consider a $G$ gauge field $a_1$ and a $\fg=Lie[G]$-valued $(d-1)$-form  gauge field $h_{d-1}$. Here we will consider the case where $h_{d-1}$ transforms under the adjoint representation of $G$. We can then construct a topological action 
\eq{
S=\itwopi \int \Tr\left[f_2\wedge h_{d-1}\right]~,
}
where $f_2$ is the field strength of $a_1$. Using this action to define a quantum theory is more subtle than the $U(1)$ case as the non-abelian gauge transformations requires one to introduce ghost fields or use BRST/BV-quantization. In this paper, we will not discuss such subtleties. 

The equations of motion
\eq{
f_2=0\quad, \quad Dh_{d-1}=0~,
}
where $D$ is the covariant exterior derivative, imply that the Wilson line $W_R=\Tr\CP\,e^{i \oint a_1}$ is topological. As mentioned in the introduction, the definition of a gauge invariant $h$-surface is subtle because the notion of path ordering, which is necessary for non-abelian gauge invariance, does not naturally extend to surface operators of higher dimension. 

In this theory we can still diagnose the possible boundary conditions. This can be done from the Lagrangian formalism either by doing canonical quantization\footnote{Here is one place where the subtlety associated to ghost fields arises. As is standard, the canonical quantization of the non-abelian gauge theory requires projecting onto gauge invariant states which requires BRST/BV quantization or the introduction of ghost fields.} or by looking at the boundary conditions from the variation of the action as above. Here we will take the approach of studying the boundary variation of the action. 

The boundary variation of the action and the gauge variation of the action are  given by 
\eq{
\delta S\big{|}_{bnd}=\int_{X_d}\Tr\left[\delta a_1\wedge h_{d-1}\right]\quad, \quad \delta_{\rm gauge}S\big{|}_{bnd}=\int_{X_d}\Tr[f_2\wedge \lambda_{d-2}]~.
}
We then see that there are two boundary conditions 
\begin{enumerate}
\item $a_1$ is fixed and flat up to gauge transformations and $D\tildeh=0$ with the constraint that  $\Tr[\Lambda \oint \tildeh_{d-1}]\in 2\pi \IZ$ where $\Lambda$ is any co-root of $G$;
\item $h_{d-1}=0$ up to gauge transformations and $a_1$ flat.
\end{enumerate}
Boundary condition 1.) is the natural Dirichlet boundary condition $|A_1\rangle$ while 2.) is naturally the Neumann boundary condition $|N\rangle$. As in the case of the $U(1)$ gauge field, this SymTFT straightforwardly accommodates flat $G$-gauge fields. However, it is unclear how to construct the analogous defect Hilbert spaces that allow for $G$-gauge fields with non-trivial characteristic classes since it is unclear how to construct the corresponding gauge invariant $\tildeh$-surfaces as discussed above. 

The Dirichlet boundary conditions clearly form an orthogonal set among the space of flat $G$-connections modulo gauge transformations as any pair of (gauge) inequivalent connections will require a non-trivial field strength in the bulk which will be projected out by the integral over $h$.  
The Neumann boundary conditions can then be constructed by summing over Dirichlet boundary conditions as the $G$-connection is free on the boundary.  

When coupling to the QFT, we can define the QFT state as above 
\eq{
\langle {\rm QFT}|=\int_{\CA_0/\CG} [dA_1]\,Z_{QFT}[A_1]~
\langle A_1|~,
}
where the path integral is over the space of flat $G$-connections $\CA_0$ modulo gauge transformations $\CG$. The Dirichlet boundary condition then exhibits the coupling of the QFT to a flat background gauge field:
\eq{\langle {\rm QFT}|A_1\rangle= \,Z_{QFT}[A_1]~.}

Additionally, this SymTFT has the capacity to encode the anomalies of $G^{(0)}$ global symmetries. This can be accomplished by introducing the corresponding Chern-Simons term
\eq{
S=\itwopi \int \Tr\left[f_2\wedge h_{d-1}\right]+i\int CS_\kappa[a_1]~,
}
where $CS_\kappa[a_1]$ is the Chern-Simons polynomial with coefficient $\kappa\in \IZ$ of the $G$-connection $a_1$. As above, this will make the Neumann boundary condition ill defined and obstructs us from gauging the $G^{(0)}$ global symmetry in the QFT. 

Because this TQFT we proposed above captures these universal features of $G^{(0)}$ global symmetries, we believe that this does indeed describe the $G^{(0)}$ SymTFT. We believe it is an interesting open problem to understand this TQFT, its operator spectrum, and categorical description in general dimension. In $d=2$ dimensional QFTs (i.e. a $2+1d$ SymTFT), this symmetry has been studied as the topological sector of $3d$ $\CN=4$ twisted $G^{(0)}$ gauge theory in \cite{Ballin:2023rmt,Costello:2018swh,Costello:2018fnz,Garner:2022vds} and directly studied in $4d$ in \cite{Cattaneo:1995tw,Cattaneo:1997eh}.

\section*{Acknowledgements}

The authors would like to thank  Ken Intriligator, John McGreevey, Po-Shen Hsin, Ibrahima Bah, Konstantinos Roumpedakis, Dan Freed, Gregory Moore, Sakura-Schafer-Nameki, Tudor Dimofte, Theo Jacobson, and especially Thomas Dumitrescu for helpful discussions. 
TDB is supported by Simons Foundation award 568420 (Simons
Investigator) and award 888994 (The Simons Collaboration on Global Categorical Symmetries). Z.S. is supported by the US Department of Energy (DOE) under cooperative research agreement DE-SC0009919, Simons Foundation award No. 568420 (K.I.), and the Simons Collaboration on Global Categorical Symmetries on the initial stage of this project.

\bibliographystyle{utphys}
\bibliography{ContSymTFTbib}

\end{document}